\documentclass[article]{aa}  
\usepackage{xcolor} 
\usepackage{graphicx}
\usepackage{lscape}
\usepackage{natbib}
\usepackage{multirow}
\usepackage{txfonts}

\newcommand{\emm}[1]{\ensuremath{#1}}
\newcommand{\emr}[1]{\emm{\mathrm{#1}}}

\newcommand{\unit}[1]{\emr{\,#1}}

\newcommand{\pcm}{\unit{cm^{-2}}} 

\newcommand{\K}{\unit{K}}
\newcommand{\kms}{\unit{km~s^{-1}}}

\newcommand{\arcs}{$^{\prime\prime}$}

\newcommand{\rev}[1]{{\bf  #1}}

\begin{document}

   \title{Molecular ion abundances in the diffuse ISM : CF$^+$, HCO$^+$, HOC$^+$, and C$_3$H$^+$. }

   \author{M. Gerin  \inst{1}
          \and H. Liszt \inst{2}
          \and  D. Neufeld \inst{3}
\and  B. Godard     \inst{1}
\and P. Sonnentrucker \inst{4,5}
\and J. Pety \inst{6}
\and E. Roueff \inst{7}
 }

   \institute{
   Sorbonne Universit\'e, Observatoire de Paris, Universit\'e PSL, 
\'Ecole normale sup\'erieure, CNRS, LERMA, F-75014, Paris, France.
  \email{maryvonne.gerin@ens.fr} %
  \and 
National Radio Astronomy Observatory, 520 Edgemont Road, Charlottesville, VA 22903, USA. 
  \and
  Department of Physics \& Astronomy, Johns Hopkins University, 3400 N. Charles St., Baltimore, MD 20218, USA.
  \and
 Space Telescope Science Institute, Baltimore, MD 21218, USA.  \and
 European Space Agency
 \and
 Institut de Radioastronomie Millim\'etrique (IRAM),
 300 rue de la Piscine, 38406 Saint Martin d'H\`eres, France.
 \and
  Sorbonne Universit\'e, Observatoire de Paris, Universit\'e PSL, 
\'Ecole normale sup\'erieure, CNRS, LERMA, F-92190, Meudon, France.
              }

   \date{Received xxx; accepted xxx}

  \abstract
{}
  {The transition between atomic and molecular hydrogen is associated with important changes
 in the structure of interstellar clouds, and marks the beginning of interstellar chemistry. 
 Most molecular ions are rapidly formed (in ion-molecule reactions) and destroyed (by dissociative recombination) in the diffuse ISM.
 Because of the relatively simple networks controlling their abundances, molecular ions  are usually good probes of the underlying 
 physical conditions including for instance the fraction of gas in molecular form or the fractional ionization.
 In this paper we focus on  three possible probes of the molecular
  hydrogen column density, HCO$^+$, HOC$^+$, and CF$^+$. }
   {We presented high sensitivity ALMA absorption data toward a sample of compact HII regions and
   bright QSOs with prominent foreground absorption, in the ground state transitions of the molecular ions HCO$^+$, HOC$^+$, and CF$^+$ and 
the neutral species HCN and HNC, and from the excited state transitions of C$_3$H$^+$(4-3) and $^{13}$CS(2-1).
  These data are compared with Herschel absorption spectra of the ground state transition of HF and p-H$_2$O.  }
   { We show that the HCO$^+$, HOC$^+$, and CF$^+$ column densities are well correlated with each other. HCO$^+$ and
   HOC$^+$ are tightly correlated  with p-H$_2$O, while they exhibit a different correlation pattern  with HF depending
  on whether the absorbing  matter is located in the Galactic disk or in the central molecular zone. 
  We report new detections of C$_3$H$^+$ confirming that this ion is ubiquitous in the diffuse matter, with an abundance relative to H$_2$ of
  $\sim 7 \times 10^{-11}$.}
{ We confirm that the  CF$^+$ abundance is lower than predicted by simple chemical models
  and propose that the rate of the main formation reaction is lower by a factor of about 3 than usually assumed.
  In the absence of CH or HF data, we recommend to use the ground state transitions of 
  HCO$^+$, CCH, and HOC$^+$ to trace diffuse molecular hydrogen, with mean abundances relative to
  H$_2$ of  $3 \times 10^{-9}$, $4 \times 10^{-8}$  and $4 \times 10^{-11}$ respectively, leading to sensitivity $N(\rm {H_2})/\int \tau dv$ of
  $4\times 10^{20}$,  $1.5\times 10^{21}$, and $6\times 10^{22}$ cm$^{-2}$/\kms \, respectively. }


   \maketitle
%

\section{Introduction}

The diffuse interstellar medium is ubiquitous in the Galaxy, and exhibits a complex structure
with a mixing of phases with widely different conditions. Dust extinction provides
the total column density of material, which can be inverted to yield
the three dimensional structure when the number of background target stars is sufficient \citep[e.g.,][]{lallement:14,green:15}.
Such recent  surveys confirm the mixing of phases with the dense structure embedded in bubbles of
low density material. However, the  extinction data do not provide information about the composition
of the absorbing gas. Classical spectroscopic tracers are commonly used for detecting emission lines from ionized gas (H$\alpha$), atomic gas (HI) or molecular gas (CO),
but they usually miss the diffuse molecular gas when the CO$(J=1 \rightarrow 0)$ emission is not strong enough to be detected.
This so-called "CO-dark  gas" because its CO emission is not detected, contributes to the total $\gamma$ ray production \citep[e.g.,][]{remy:17}, 
dust extinction or emission \citep{abergel:11}, and [CII] fine structure line emission \citep{pineda:13}. 

The CO-dark gas usually surrounds the molecular clouds detected in CO \citep{wolfire:10,remy:17}. Indeed,
molecular clouds are known to occupy a small fraction of the Galaxy volume,
 although they represent most of the dense gas mass. The transition region
 between fully molecular and fully atomic gas extends over larger areas on the
 sky  and  accounts for a significant, although not dominant, fraction of the neutral gas mass
at the Galaxy scale, $\sim 30\%$ \citep{pineda:13}.
  These authors have used {\sl Herschel} data and performed a careful separation
 of the contributions of the neutral and molecular ISM phases by comparing the
 [CII] emission with that of HI and CO. To derive column densities from the emission line
 intensities, these authors have made standard  assumptions on the
 HI spin  temperature, and on the scaling of CO ($J=1\rightarrow 0$) emission with the H$_2$
 column density. Both methods have their own uncertainties which propagate
in the derivation of the fraction of molecular gas not traced by CO.
 Alternative tracers are therefore needed which can probe the diffuse gas, 
be easily accessible from the ground, and provide an independent estimation
of the diffuse molecular gas properties.

Absorption spectroscopy provides an
alternative means for separating the ISM phases. In low density regions, the excitation of molecular energy
levels is dominated by the cosmic background radiation. Therefore the derivation 
of line opacities and molecular column densities is straightforward and accurate. As summarized by \citet{snow:06} UV and visible spectroscopy have been extensively 
used to probe the properties of the diffuse and translucent gas, understand the mechanisms that control the transition from atomic to molecular gas and investigate
 the composition and physical conditions of the diffuse molecular gas  including the CO-dark regions \citep[e.g.,][]{spitzer:75,rachford:09,jenkins:11,sheffer:08}.
Similarly, absorption spectroscopy has been successfully used to probe the structure and chemistry of the diffuse interstellar
medium at radio wavelengths for decades \citep{goss:68,whiteoak:70,lucas:96}.
The ground state line of hydrides observed with the {\sl Herschel} satellite provided a wealth of new informations on the properties of the diffuse
interstellar gas \citep{gerin:16}, including  sensitive 
tracers of molecular gas (HF, CH, and H$_2$O), of the cosmic ray ionization
rate (OH$^+$, H$_2$O$^+$, H$_3$O$^+$), and strong
supporting evidence for the role of turbulent dissipation in
triggering the formation of the first molecule building blocks like CH$^+$ and
SH$^+$ \citep{godard:12,godard:14}. The GREAT receiver on SOFIA has enabled further studies
of CH, OH, and SH absorption \citep{wiesemeyer:16,wiesemeyer:17,neufeld:15}. 

Among the interstellar molecules, fluorine species are particularly attractive molecular diagnostics because of the simplicity of 
fluorine chemistry. Fluorine atoms are unique because they can react exothermically with
H$_2$ to form HF, the only neutral diatomic hydride which is more
strongly bound than H$_2$. As a consequence theoretical models predict
that HF will become the dominant fluorine reservoir in molecular
gas, a prediction fully confirmed by the {\sl Herschel} observations \citep{sonnentrucker:15}. HF is
destroyed mainly by reactions with C$^+$, forming CF$^+$. Hence,
CF$^+$ is expected to be the second most important fluorine reservoir
in regions where C$^+$ is abundant, and it is expected to account for
a few percents of the gas-phase fluorine abundance. First detected
in the Orion Bar \citep{neufeld:06}, CF$^+$ has been shown to be present  in the diffuse ISM \citep{liszt:14,liszt:15},
with a somewhat lower abundance (by a factor of $\sim 4$) than the theoretical predictions. 

HCO$^+$ and HOC$^+$ are two interesting ions that are widespread in the diffuse interstellar medium.
While  HCO$^+$, the most stable isomer, has been shown from observations to be a good tracer of
molecular hydrogen since 20 years \citep{lucas:96}, the absorption becomes saturated at 
H$_2$ column densities corresponding to translucent gas, with visual extinction between 1.5 and 3 mag \citep{snow:06}. It is therefore interesting to look for
complementary tracers with different sensitivities to the molecular hydrogen column allowing to probe a broader range of H$_2$ column densities.
The less stable isomer, HOC$^+$ may be such an alternative tracer. With an abundance ratio to HCO$^+$ at the
percent level \citep{liszt:04}, HOC$^+$ is thought to be 
mainly formed by the chemical reaction of C$^+$ with H$_2$O, whose secondary channel
leads to HCO$^+$.  Therefore  HCO$^+$ and HOC$^+$  are expected to coexist in the diffuse molecular gas.

Another interesting property of CF$^+$ and HOC$^+$ is their expected  close relationship to their parent species, C$^+$ and HF for CF$^+$,
and C$^+$ and H$_2$O for HOC$^+$. At submillimeter
wavelengths, Galactic HF and H$_2$O cannot be detected from the ground because of the high opacity of the atmosphere at the
frequencies of their ground state rotational transitions. Also, the ground electronic transitions for HF and H$_2$O occur in the FUV around 950\AA \, for HF 
and 1240\AA \, for H$_2$O but the combination 
of weak band strength and blending with adjacent atomic species renders their detection and study very challenging even from space.  
The ground state transitions of the three molecular ions discussed here are located at millimeter 
wavelengths where the atmospheric transmission is excellent. We have therefore
performed observations with ALMA of sources previously observed
with {\sl Herschel} to probe how the ions and neutral species are related and whether the HCO$^+$, HOC$^+$ and CF$^+$ molecular ions could be
used as surrogates for either the hydride species or molecular hydrogen.

The observations are described in Section \ref{sec:obs} and analyzed in Section \ref{sec:analyse}. Section \ref{sec:abundances} presents
the resulting abundances relative to H$_2$ and the implication for the chemistry of diffuse gas are discussed in Section \ref{sec:chemistry}.
Section 6 discusses the ways in which gas near the Galactic center differs from that in the disk.
Section \ref{sec:conclusion} summarizes the main conclusions.
  
\section{ALMA observations}
\label{sec:obs}

We present ALMA Cycle 2 observations under project 2013.1.1194.S: The targeted sources are listed in 
Table \ref{tab:sources}. Some of the observations under this project were previously presented by 
\cite{gerin:17} who discussed  absorption from gas in the Galactic bulge that is seen toward  the quasar
background target B1741-312.

 Observations were  divided in three scheduling blocks that have been observed
separately in different interferometer configurations between 2014 and 2015.
We used a configuration of the ALMA correlator allowing simultaneous observations of the lines listed in Table \ref{tab:lines} using
the two available sidebands, with spectral resolution set to 61.03 kHz  (0.18 \kms) for CF$^+$, 122.07 kHz (0.4 \kms) for HOC$^+$, and 244.14 kHz (0.8 \kms) 
for HCN,  HNC,  HCO$^+$, c-C$_3$H, $^{13}$CS, and C$_3$H$^+$.
A 1.875~GHz wide spectral window with a coarse spectral resolution of 15.6~MHz  was reserved for detecting the continuum emission at 102 GHz with high sensitivity.
The bandpass calibrators were J1427-4206, J1717-3342, J1924-2914, and the phase calibrator was J1752-2956. 
As flux calibrator the solar system objects Titan, Venus, and Ceres
were used. The system attempted with success to use B1730-130 aka J7133-1304, one of the program targets,
 as bandpass calibrator and data affected by this problem were  masked.
The data were calibrated and deconvolved using CASA and the resulting data-cubes were then exported
 to GILDAS for further analysis. 

Table \ref{tab:sources} lists the continuum values at 102 GHz derived from the
 spectral window dedicated to the detection of the continuum. The 102 GHz continuum window was used only for phase calibration.  The continuum level to be subtracted to measure the strength of the absorption was determined separately in each spectral window.
 Images of the four extended sources (G0.02-00.7, SgrB2, G5.89-0.4, G10.62-0.4) are shown in Fig \ref{fig:app:images}. 
  Absorption spectra have been extracted at the peak of the continuum maps or using weighted average of bright pixels
 with minimum line contamination from the background source. The positions of the extracted spectra are listed in Table \ref{tab:sources}. 
  The absorption spectra have been obtained by  dividing the observed spectra by the continuum intensity in the same band. For 
  further analysis the data have been smoothed at a spectral resolution of 1~\kms. 
  Except for the very weak sources in the G0.02-00.7 field, we obtain a noise level between 0.2 and 0.6 \% of the continuum brightness
 when smoothed at the common velocity resolution of 1~\kms, allowing the detection of very weak features.

\begin{figure*}
    \includegraphics[width=0.8\textwidth]{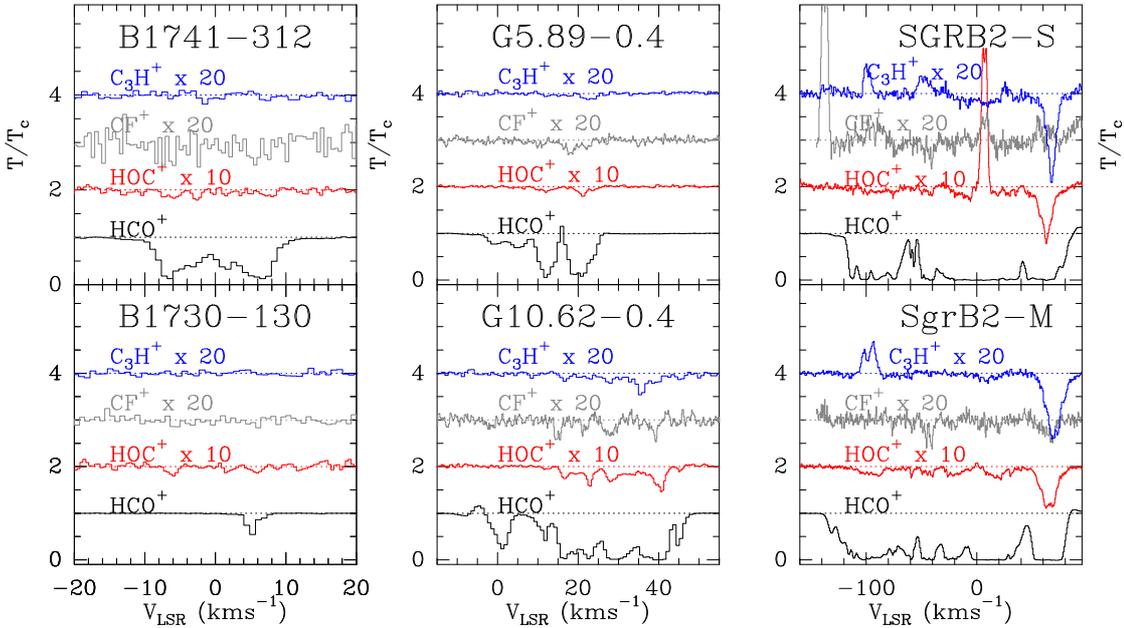}
    \caption{Presentation of the molecular ion spectra toward the target sources at the peak of the continuum emission. The data are displayed as line to continuum 
 ratio, and shifted vertically for clarity. The weak absorption lines have been multiplied by the indicated scaling factors: 
20 for C$_3$H$^+$ and  CF$^+$ and 10 for HOC$^+$. 
} 
              \label{fig:spec}%
    \end{figure*}

 Figure \ref{fig:spec} presents the ion spectra, and  Figure \ref{fig:app-spec} shows the whole set of spectra. 
 The continuum emission of the compact HII regions in the field of G0.02-0.07 are too weak for detecting weak absorption
 features produced by HOC$^+$ and CF$^+$. Only the strongest features produced by HCO$^+$, HCN, and HNC have been detected. 
The data toward the compact sources in the G0.02-0.07 field will not be further discussed.

\begin{table*}
\caption{Summary of observations}
\label{tab:sources}
\begin{tabular}{lcccccccc}
\hline
\hline
Source$^a$ & RA$^a$ & Dec$^a$ & RA$_s$$^b$ & Dec$_s$$^b$ & V$_{LSR}$$^c$ & Beam & $F_{s}$$^d$  & $\sigma$$^e$\\
       & (J2000)   & (J2000)    & (J2000) & (J2000) & km/s & \arcs & Jy/beam & \% \\
\hline
B1730-130 & 17:33:02 & -13:04:49 &  17:33:02.7 & -13:04:49 &                      ...  &  $3.7 \times 1.7$ & 1.9 & 0.32\\
\hline
G5.89-0.4 & 18:00:30 & -24:04:00 & 18:00:30.4 & -24:04:01.3 &                    9.0 & $1.0 \times 0.6$ & 0.34 & 0.28 \\
G10.62-0.4 & 18:10:28 & -19:55:50 &    18:10:28.7 & -19:55:50 &                       -0.4 & $0.9 \times 0.6 $& 0.46 & 0.3 \\
\hline
SgrB2-M & 17:47:20 & -28:23:45 &  17:47:20.3 & -28:23:05 &                             60 & $2.7 \times 1.8$ & 2.8 & 0.4\\
SgrB2-S & 17:47:20 & -28:23:45 & 17:47:20.4 & -28:23:45 &                                 60 & $2.7 \times 1.8$ & 0.58 & 0.7\\
G0.02-0.07$^f$ & 17:45:51 & -28:59:50 &  17:45:51.9 & -28:59:27 &                     50 & $2.7 \times 1.8$ & 0.02 & 6.3\\
                    &    17:45:51 & -28:59:50     &      17:45:52.1 & -28:59:41 &         50               &     $2.7 \times 1.8$                      & 0.008 & 15 \\
                    &.   17:45:51 & -28:59:50       &       17:45:52.4 & -29:00:03 &       50                 &  $2.7 \times 1.8$                         & 0.01  & 13 \\
                    &    17:45:51 & -28:59:50         &       17:45:51.6 & -29:00:22 &      50                  &  $2.7 \times 1.8$                        & 0.03 & 13 \\
B1741-312 & 17:44:23 & -31:16:35 &   17:44:23.6 & -31:16:36 &                    ...  & $2.6 \times 1.9$ & 0.55 & 0.6\\
\hline
\end{tabular}
\tablefoot{$^a$ Field center. $^b$ Position of the extracted spectrum. $^c$ LSR Velocity of the molecular gas associated with the source. $^d$ Continuum value at 102~GHz at the position of the extracted spectrum.  $^e$ mean noise level on the continuum  at 1\kms spectral resolution, expressed as percentage of the continuum intensity. $^f$ G0.02-0.07 is part of the  SgrA molecular complex. The sources are displayed according to the completion of observations. 
}
\end{table*}

Toward both G5.89-0.4  and G10.62-0.4, which have ultra-compact HII regions, the CF$^+$ data are
contaminated by the emission of the H66$\epsilon$ hydrogen recombination line, which was already seen in
the W49N analysis of \citet{liszt:15}. We chose a position away from the brightest continuum and fitted the
broad hydrogen recombination line with a Gaussian profile. Other hydrogen recombination lines are
detected, notably the H59$\gamma$ at 89.1985450~GHz which is blue-shifted by 10 MHz (33 \kms) from the HCO$^+$(1-0) transition.  
The hydrogen recombination line emission is associated with the bright free-free continuum emission, and
is absent in the QSO spectra. Because H66$\epsilon$ is redshifted relative to CF$^+$, it does not
contaminate the data toward SgrB2 where the line of sight absorption is blue-shifted relative
to the main envelope. 

A faint and broad absorption from C$_3$H$^+$ is detected at velocities between 15 and 45~kms$^{-1}$ corresponding to the diffuse gas along the
line of sight to the G10.6-0.4 HII region, with a superimposed narrow feature at 35~kms$^{-1}$. A second narrow feature is present outside the expected spectral
range of the foreground absorption near $-167$~kms$^{-1}$ (not displayed here).  While these narrow absorption lines are real, they are not produced
by C$_3$H$^+$, but are associated with molecular lines from the hot molecular core next to G10.6-0.4, because we detect weak emission at the
same frequency a few pixels away from the bright continuum emission.
Similar narrow absorption features  are detected towards G10.62-0.4 in the HCN and HOC$^+$ spectral windows, notably a rather strong feature at 124~kms$^{-1}$, not displayed here.
 We used Splatalogue \footnote{www.cv.nrao.edu/php/splat/advanced.php} and the CDMS\footnote{www.astro.uni-koeln.de/cdms} and JPL\footnote{spec.jpl.nasa.gov} line catalogs for line identification and assignment.
We assigned the narrow feature appearing at 124~kms$^{-1}$ in the HCN spectral window  to the CH$_3$OH 15(3,13)-14(4,10) line at 88594.4809 MHz, leading to
a LSR velocity of -1.2 \kms  for this narrow absorption feature. Then, the two narrow features detected in the C$_3$H$^+$ spectrum can be assigned
to two transitions of g-CH$_3$CH$_2$OH, the 10(0,10)-10(1,10) line at 89947.016 MHz ($E_l = 70.98$~cm$^{-1}$) for the 35~kms$^{-1}$ feature and the
13(1,12)-13(1,13) line at 90007.685 MHz ($E_l= 90.9$~cm$^{-1}$) for the $-167$~kms$^{-1}$ feature. 
The spectroscopy of CH$_3$CH$_2$OH is well known, because it has been studied by \citet{Pearson:08} and revised by \citet{muller:16}.
We have fit the narrow absorption lines with Gaussian profiles, and used the residual to estimate the absorption due to C$_3$H$^+$ along the G10.62-0.4 line of sight.

\begin{table*}
\label{tab:lines}
\caption{Observed lines}
\begin{tabular}{lcrrr}
\hline
Species        & Transition & Frequency & A                           &                          N/$\int \tau dv$ $^a$ \\
                    &                  &     MHz       & 10$^{-5}$ s$^{-1}$ & 10$^{12}$ cm$^{-2}$/kms$^{-1}$ \\
   \hline
HCO$^+$    	&    1 -- 0     		& 89188.525  		& 4.19                    	& 1.11 \\
HOC$^+$  	&    1 -- 0           	& 89487.414  		& 2.13   			&  2.19 \\ 
CF$^+$$^b$ 		&    1 -- 0           	& 102587.533 	&  0.48  		&  13.5 \\ 
C$_3$H$^+$	&    4 -- 3           	& 89957.625 		& 3.39   			&   16.2 \\ 
c-C$_3$H 	& $2(1,2) 5/2 , 3 - 1(1,1) 3/2, 2$ &  91494.349     &   1.55            & 14.3 \\
$^{13}$CS 	&  2 --1 			& 92494. 257 		& 1.18			&	9.72	\\
HCN$^b$ 		& 1 -- 0 			&   88631.601 		&  2.41	                 & 	1.91	\\
HNC 		& 1 -- 0 			& 90663.564 		& 2.69			& 	1.78	\\
NH$_2$CHO$^b$ 	& $2(1,2)-1(0,1)$ 	& 102217.572 		& 0.27		& 	114	\\
\hline
\end{tabular}
\tablefoot{$^a$ For an excitation temperature equal to the CMB, 2.725 K, $^b$ for the combined hyperfine components. A recent determination of the rest frequency has been published by \citet{stoffels:16}}.
\end{table*}

The column densities have been derived by integrating the line opacity over selected velocity intervals, either 1\kms \, for the comparison of the line profiles or broader intervals as listed in 
Table \ref{tab:column} for deriving column densities associated with individual velocity features. Because the analysis is focused on the diffuse
gas, we have assumed that the level population is in equilibrium with the Cosmic Microwave Background. This hypothesis is justified by the low density of the medium as discussed for instance by 
\citet{liszt:16} for the G10.62-0.4 sight-line.  We recall the expression for the coefficient $N/\int \tau dv$ for a ground state transition :
\begin{equation}
N = Q(T_{ex}) \frac{8\pi \nu^3}{c^3} \frac{1}{g_uA_{ul}} \Big[ 1-e^{-\frac{h\nu}{k_BT_{ex}}} \Big]^{-1} \int \tau dv
\end{equation}
where $T_{ex}$ is the excitation temperature, $Q(T_{ex})$ the partition function, $A_{ul}$ the Einstein coefficient of the transition of frequency $\nu$
between the upper and lower levels of degeneracies $g_u$ and $g_l$. The coefficients we have used are listed in the last column of Tab. \ref{tab:lines}.

\section{Results}
\label{sec:analyse}

 %

\subsection{HOC$^+$ and HCO$^+$}
\label{sec:hocp}

\begin{figure}
\includegraphics[width=0.45\textwidth]{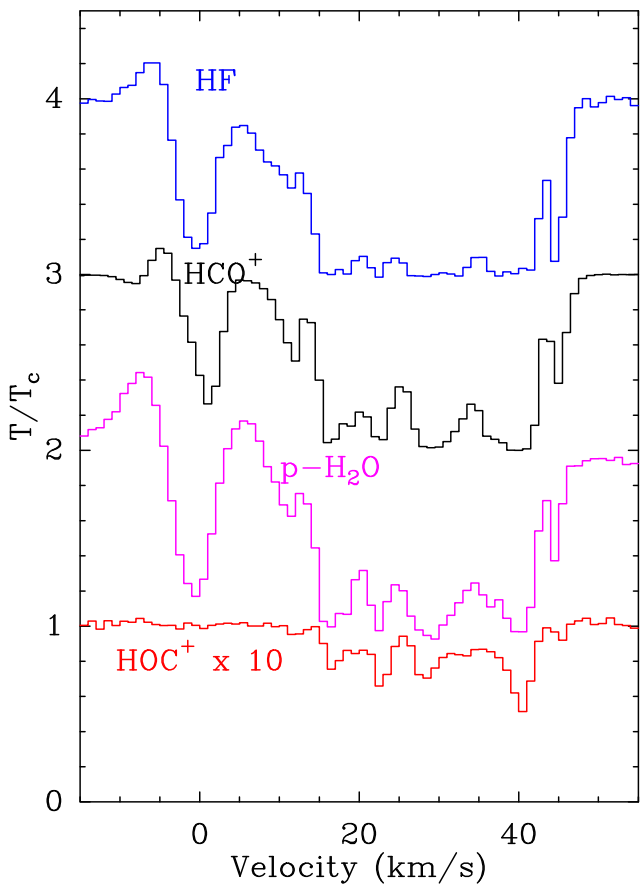}
\caption{Comparison of absorption spectra toward G10.62-0.4 obtained with {\sl Herschel} (p-H$_2$O and HF) and with ALMA (HCO$^+$ and HOC$^+$)
at the same velocity resolution of 1 \kms. The spectra have been displaced vertically for clarity. The HOC$^+$ spectrum is  scaled by a factor of 10. The velocity range between $-10$ and 10~kms$^{-1}$ 
corresponds to the molecular gas associated to the G10.6-0.4 HII region, while the foreground absorption can be seen between  10  and 50~kms$^{-1}$.
 Despite the huge difference in angular resolution, the absorption spectra  are very similar.}
\label{fig:h2o}
\end{figure}

\begin{figure}
\includegraphics[width=0.45\textwidth]{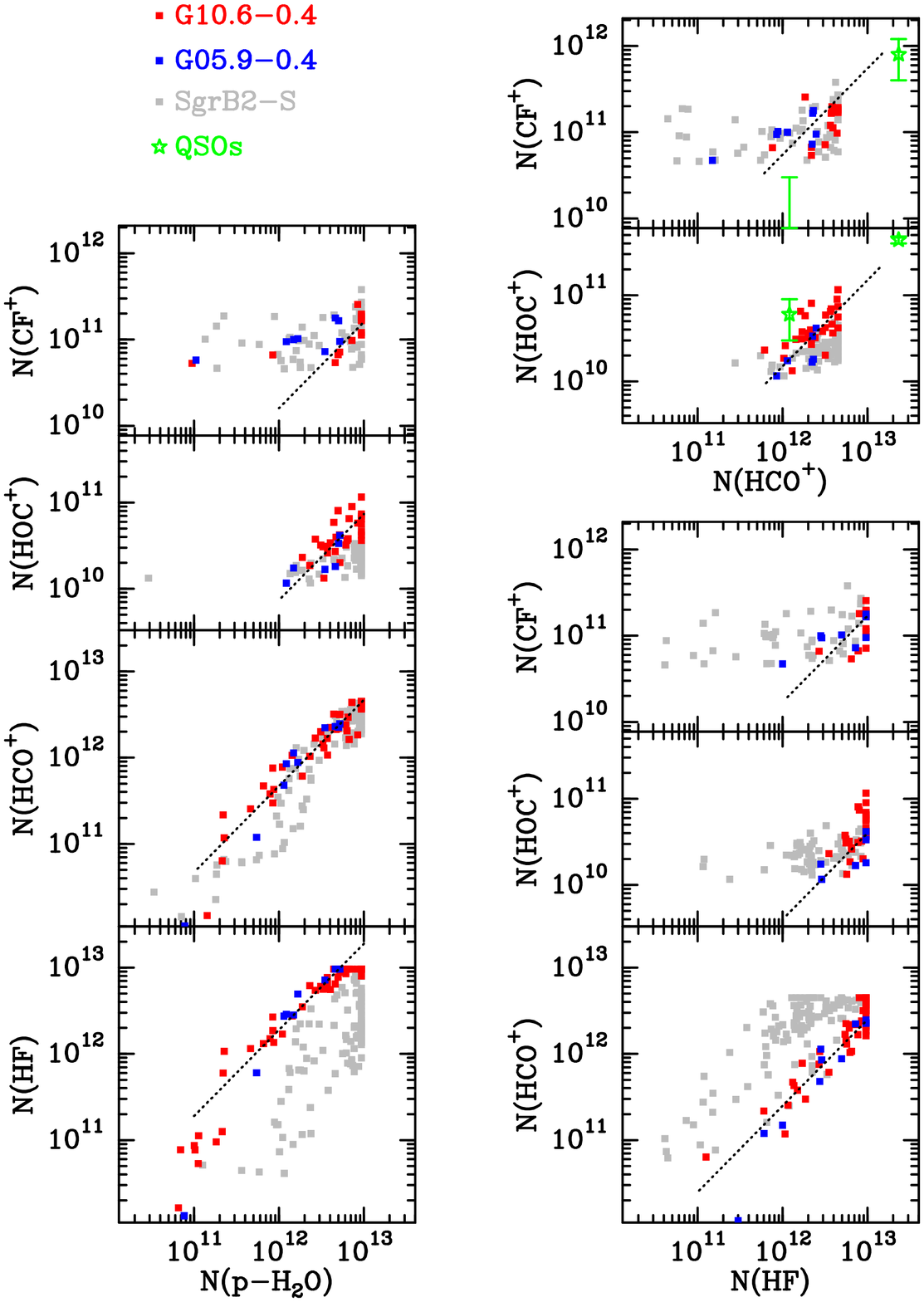}
\caption{ Comparison of the column densities in 1 kms$^{-1}$ velocity bins for the program sources. Data for SgrB2-S are plotted in gray, G10.62-0.4
in red, G5.89-0.4 in blue and those toward B1730-130 and B1741-312 in green with error bars.  The dotted lines show the mean relationships discussed in the text. The HF and p-H$_2$O data have been obtained with {\sl Herschel} \citep{sonnentrucker:15}, the other data are
from ALMA. The velocity range corresponding to the background
sources are excluded. Only data points with SNR greater than $ 2.5 \sigma$, and with opacity lower than $3.5$ are shown.  }
\label{fig:correl}
\end{figure}

As illustrated in Fig. \ref{fig:spec} and Fig. \ref{fig:h2o}, the profiles of HCO$^+$ and HOC$^+$ are similar, confirming that these species
are closely related to each other in diffuse gas. We have compared the absorption profiles of both ions with those of p-H$_2$O and HF
detected by {\sl Herschel} \citep{sonnentrucker:15}. Figure \ref{fig:h2o} presents a comparison of the line profiles for the case of G10.62-0.4, 
while Figure \ref{fig:correl} presents   channel by channel comparisons for all sources. Each point in this figure represents a 1\kms \, velocity channel,
the red points correspond to the G10.6-0.4 line of sight, the blue points to G5.89-0.4, the green points to the QSO sight-lines and the 
gray points to SgrB2-S. This plot only shows data corresponding to the material along the line of sight, namely velocities higher than 8~\kms \, for G10.62-0.4, 15.5~\kms \, for G5.89-0.4 and lower than $-20$~\kms \, for SgrB2-S. We have not applied any restriction for the QSO sight-lines.
It is remarkable that, when comparing HCO$^+$  with p-H$_2$O, all sources lay on the same relationships, while
 SgrB2-S stands out when HF or CF$^+$ are compared with HCO$^+$ or with p-H$_2$O. When compared with p-H$_2$O or with
 HCO$^+$, the HOC$^+$ absorption toward SgrB2-S is somewhat fainter
 than toward the sources in the Galactic plane, but the difference is not as strong as with HF.  Differences between
the Galactic center and Galactic disk gases are discussed in more detail in Section 6.
 
The relationships between p-H$_2$O and HCO$^+$ and  p-H$_2$O and HOC$^+$, like that  between HCO$^+$ and HOC$^+$, are nearly linear.
Because of the different behavior of the gas in the Galactic Center, we have restricted the correlation analysis to the G10.62-0.4 and G5.89-0.4 sight-lines. This will also allow us to
use a consistent sample for all species, those observed with {\sl Herschel} like p-H$_2$O and HF, and those observed with ALMA. 
We obtained the following relationships from  least square fits, assuming the column densities are proportional to each other and restraining the fit to points with SNR greater than 2.5$\sigma$ and 
opacity lower than 3.5 :
\begin{equation}
N({\rm HCO^+}) = (0.47  \pm 0.03) \, N({\rm p-H_2O})
\end{equation}

\begin{equation}
N({\rm HOC}^+) = (7.4 \pm 1.5) \times 10^{-3}  \, N({\rm p-H_2O})
\end{equation}

\begin{equation}
N({\rm HOC}^+) =  (1.5  \pm 0.3) \times 10^{-2}  \,  N({\rm HCO^+})
\end{equation}

The excellent correlation between HCO$^+$ and H$_2$O was already discussed in the context of the SWAS and ODIN missions,
using the ground state o-H$_2$O line at 557~GHz \citep{neufeld:02,plume:04}, while \cite{lucas:96} showed that HCO$^+$ and OH 
absorption were tightly correlated. A tight correlation between HCO$^+$ and H$_2$O is therefore expected, and clearly seen in the data.

As for HOC$^+$,  \citet{liszt:04} observed HCO$^+$ and HOC$^+$ toward three QSOs with the Plateau de Bure
Interferometer (PdBI) and derived a mean ratio of N(HOC$^+$)/N(HCO$^+$) $\sim 0.01$, similar to the ALMA measurement, although with larger uncertainties. 
The new detections obtained
with ALMA toward B1741-312 and B1730-130 \rev{follow  the same relationship as previous measurements obtained with PdBI and allow a more precise determination 
of the N(HOC$^+$)/N(HCO$^+$) abundance ratio of $(1.5  \pm 0.3) \times 10^{-2}$}.
\cite{liszt:04} derived $N($HOC$^+$$)/N($OH$) =  3 - 6 \times 10^{-4}$ by comparing the Plateau de Bure data with absorption
measurements of the OH $\Lambda$ doubling lines at 18~cm. The recent analysis of SOFIA and {\sl Herschel}
observations performed by  \citet{wiesemeyer:16} provides an independent means to relate the OH and H$_2$O column densities.  
These authors obtained  $N($OH$)/N($p-H$_2$O$) \sim 16$ or $N($OH$)/N($H$_2$O$)  \sim 4$
using an ortho-to-para ratio of 3 for H$_2$O \citep{flagey:13}. 
The new ALMA measurements therefore predict N(HOC$^+$)/N(OH) $\sim 4.5 \times 10^{-4}$, in  good agreement
with the independent determination toward distant QSOs.

We have also compared HCO$^+$ and HOC$^+$ with HF. As illustrated in Fig. \ref{fig:correl}, the velocity channels belonging to
the SgrB2-S sightline lead to different relationships than those toward G10.6-0.4 and G5.89-0.4.  A similar behavior
has been noticed by \citet{sonnentrucker:13} for HF and p-H$_2$O, where an excess of water vapor absorption relative
to HF was detected for the molecular gas in the vicinity of the Galactic Center. The excess of HCO$^+$ and HOC$^+$
absorption relative to HF for gas in the Galactic Center appears therefore as another consequence of the tight relationships between p-H$_2$O and HCO$^+$, and to a lesser extend because of the limited number of velocity channels where both species are detected, between p-H$_2$O and  HOC$^+$.
We obtained the mean relationships for Galactic disk diffuse gas :

 \begin{equation}
N({\rm HCO}^+) = (0.25 \pm 0.05)  \, N({\rm HF})
\end{equation}

\begin{equation}
N({\rm HOC}^+) =  (4  \pm 1.5) \times 10^{-3}  \,  N({\rm HF})
\end{equation}

\subsection{CF$^+$}
\label{sec:cfp}

Because the CF$^+$ absorption is very weak, while the HF lines are heavily saturated, there are very few velocity channels for which 
CF$^+$ and HF can be compared, leading to a relatively high uncertainty on the abundance ratio. We  also show a comparison of the CF$^+$ column density with 
 that of p-H$_2$O and of HCO$^+$ 
in Figure \ref{fig:correl}. A first comparison of CF$^+$ and HF based on NOEMA and {\sl Herschel} observations of the line of sight toward the
massive star forming region W49 has been
presented by \cite{liszt:15}.  They found that CF$^+$ is well correlated with HF as expected but that the
mean ratio is  lower by a factor of $\sim 4$  than the predicted value of $\sim 1/20$  using the reaction rates by \citet{nw:09} listed in Table \ref{tab:rates}.
Our results obtained using the G10.62-0.4 and G5.89-0.4 sight-lines support the previous measurements with

\begin{equation}
N({\rm CF^+}) =  (1.7 \pm 0.3) \times 10^{-2}  \, N({\rm HF})
\end{equation}

We also obtain the following mean relationships with HCO$^+$, HOC$^+$ and p-H$_2$O :

\begin{equation}
N({\rm CF^+})  = (5.5 \pm 1) \times 10^{-2} \, N({\rm HCO^+})
\end{equation}

\begin{equation}
N({\rm CF^+})  = (2.5 \pm 0.7)  \, N({\rm HOC^+})
\end{equation}

\begin{equation}
N({\rm CF^+})  = (1.6 \pm 0.5 ) \times 10^{-2} \, N({\rm p-H_2O})
\end{equation}

\subsection{C$_3$H$^+$}
\label{sec:c3hp}
The molecular ion C$_3$H$^+$ has been first identified in the ISM as the carrier of a series of unidentified lines in the Horsehead nebula
WHISPER line survey \citep{pety:12}, and then detected in PDRs like the Orion Bar \citep{cuadrado:15} and along the line of sight
to SgrB2 \citep{mcguire:13}. \citet{guzman:15} have imaged its spatial distribution at the 
Horsehead nebula edge and conclude that C$_3$H$^+$ behaves as a precursor of hydrocarbons, and that its chemistry is related to the destruction
of carbonaceous grains and PAHs.  We detect a weak and broad absorption in the velocity range corresponding to the ISM along the
sightline toward G10.62-0.4 as well as additional absorption features toward G5.89-0.4 and in the SgrB2 envelope near 60 kms$^{-1}$ (Fig \ref{fig:spec}).
The achieved S/N ratio is not sufficient for the QSO sight-lines, and
the spectra toward our two SgrB2 targets are too crowded with multiple emission lines and possible spectral confusion with low excitation lines from CH$_3$NH$_2$ for a derivation of C$_3$H$^+$ column densities in these sources. The derived column densities reported in Table \ref{tab:column} are at the level of a few times 10$^{11}$ cm$^{-2}$, very similar
to the values found at the edge of the Horsehead nebula.
C$_3$H$^+$ is then about two times more abundant than HOC$^+$, with a mean abundance ratio relative to HCO$^+$ of $\sim 2.5\%$. 

\section{Molecular ion abundances}
\label{sec:abundances}
\begin{figure}
\includegraphics[width=0.48\textwidth]{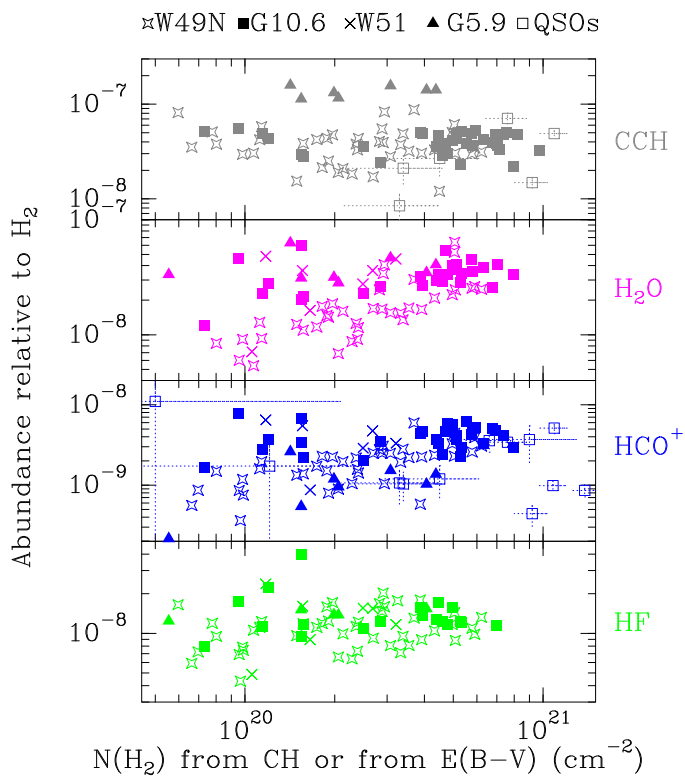}
\caption{ Abundance relative to H$_2$ as a function of the H$_2$ column density derived from the submillimeter CH observations at 532 or 536 ~GHz or from the reddening.
The different symbols refer to different sight-lines as noted at top. HCO$^+$ is shown in blue, HF in green, H$_2$O in 
purple and CCH in gray. The CCH data  were taken with the IRAM-30m telescope \citep{gerin:11} as well as the HCO$^+$  data for W49N and W51 \citep{godard:10}.}
\label{fig:abond}
\end{figure}

We report in Table \ref{tab:column} the column densities of the detected species in selected velocity intervals. We have also
included column densities for the detected neutral species. The H$_2$O abundance refers to both ortho and para symmetry states.
We first discuss the HCO$^+$ abundance relative to H$_2$, using the {\sl Herschel} CH absorption measurements \citep{gerin:10} to derive the
H$_2$ column densities, and the well determined CH abundance relative to H$_2$, [CH] = $3.6 \times 10^{-8}$ \citep{sheffer:08}.
 Figure \ref{fig:abond} presents  the relative abundances of HCO$^+$, HF, p-H$_2$O and CCH as a function of the H$_2$ column density.
 For the QSO sight-lines, without CH data, the H$_2$ column density has been derived using
the mean reddening E(B-V) from \cite{schlegel:98} and with a correction for the atomic gas using HI absorption \citep{lucas:96}.
The {\sl Herschel} and ALMA data are complemented by absorption data of HCO$^+$ and CCH using the IRAM-30m telescope  \citep{godard:10,gerin:11}. Including the sight-lines towards 
W49N and W51 increases the sample size for a more robust statistics. These sight-lines have been shown to probe similar environments as those towards G10.62-0.4 and G5.89-0.4 \citep{gerin:15}.  In addition to the species discussed in this paper (HCO$^+$, HF, and H$_2$O), we have included the CCH radical because it is closely chemically related to CH and as such is a potential tracer of  molecular hydrogen in
diffuse gas. 
Over a range of column densities spanning more than one order of magnitude, and for all sight-lines probing
different environments, we confirm that the HCO$^+$ abundance relative to H$_2$ remains fairly
constant with a mean value of [HCO$^+$] = $3.1 \times 10^{-9}$ and a moderate scatter of 0.21 dex. HF and CCH also present nearly constant abundances relative to H$_2$ with
moderate dispersions of about 0.15 dex. 

Using the HCO$^+$ abundance as reference, with [HCO$^+$] = $ 3\times 10^{-9}$, we list in Table \ref{tab:abundances} the abundances relative to H$_2$ of the HOC$^+$, 
CF$^+$, and C$_3$H$^+$ molecular ions. We also recall the mean HF, CCH, H$_2$O and OH abundances for the same sight-lines.
The abundance of HF relative to H$_2$ has been directly determined toward the background star HD154368 by \citet{indriolo:13}
to be $(1.15 \pm 0.4) \times 10^{-8}$ in agreement with the mean value listed in Table \ref{tab:abundances}. This sight-line has
$N(HF)/N(CH) = 0.26 \pm 0.06$, close to the mean value determined in the far infrared with {\sl Herschel}, 0.4 \citep{godard:12, wiesemeyer:16}.
 
The new ALMA data confirm previous measurements and expand the range of probed physical conditions. It is remarkable that the diffuse
molecular gas, detected locally toward background QSOs, or in the Galactic Plane along sight-lines
toward distant HII regions, presents rather uniform properties. The rare molecular ions CF$^+$, HOC$^+$ and C$_3$H$^+$
reach abundances relative to H$_2$ of a few times 10$^{-11}$. The derived value for C$_3$H$^+$ of $\sim 7 \times 10^{-11}$ is in  
  good agreement with the abundance measured at the edge of the Horsehead nebula or the Orion Bar from C$_3$H$^+$ emission lines \citep{pety:12,cuadrado:15,guzman:15}.
  
The homogeneity of the relative molecular abundances can be used to select the best diagnostics for detecting diffuse molecular
gas in absorption. The last column of Table \ref{tab:abundances} lists the detectable H$_2$ column for an integrated  opacity of the
ground state transition of 1~\kms \,using the mean H$_2$ abundances listed in column 2 of the same Table. For comparison,
the sensitivity of the CH ground state transitions is $N($H$_2$$)/\int \tau dv$ = 10$^{21}$ cm$^{-2}$/\kms \, for the 532 and 536 GHz lines 
and $9.7 \times 10^{20}$  cm$^{-2}$/\kms \, for the stronger transitions near 2.0 THz now accessible
with SOFIA \citep{wiesemeyer:17}.
We see that HF is the most sensitive
tracer of diffuse molecular hydrogen, and that OH, p-H$_2$O, and HCO$^+$ have comparable sensitivities of 
$N($H$_2) \sim 3 - 4 \times 10^{20}$ \pcm \, for
an integrated absorption of 1~\kms. For larger gas columns, when
the ground state transitions from this first set of species become saturated, CCH and CH, and then HOC$^+$ can be used
as probes of diffuse molecular gas. The millimeter lines from CF$^+$ and C$_3$H$^+$ appear too weak to be used as
diagnostics of diffuse molecular gas.

\begin{table*}
\label{tab:column}
\caption{Column densities}
\begin{tabular}{lcccccccccc}
\hline
Source & $N($HOC$^+$$)$       & $N($CF$^+$$)$            & $N($C$_3$H$^+$$)$    & $N($HCO$^+$$)$    & $N($c-C$_3$H$)$ & $N( $HNC$)$ & $N($$^{13}$CS$)$ \\
  \kms          & 10$^{12}$  \pcm   & 10$^{12}$  \pcm    & 10$^{12}$  \pcm    & 10$^{12}$  \pcm   &10$^{12}$  \pcm   & 10$^{12}$  \pcm   & 10$^{12}$  \pcm      \\
\hline
G10.62-0.4 &                      &                              &                               &                          &                         &               &                     \\ 
$[7 , 21]$   & $0.26 \pm 0.02$	  & $0.54 \pm 0.15$   & $0.38 \pm 0.16$    & $16.7 \pm 0.5$  & $1.7 \pm 0.4$  & $6.4 \pm 0.9$    & $0.39 \pm 0.1$\\
$[21 , 26]$ & $0.22 \pm 0.01$ & $ 1.0 \pm 0.1$      & $0.25 \pm 0.10$      & $10.5 \pm 0.3$   & $1.4 \pm 0.4$  & $5.9 \pm 0.9$    &$0.21 \pm 0.05$\\
$[26 , 35]$& $0.39 \pm 0.02$  & $ 0.53 \pm 0.12$  & $0.90 \pm 0.16$    & $26.6 \pm 0.7$  &$0.9\pm 0.4$       & $14 \pm 1$        &$0.52 \pm 0.06$\\
$[35 , 43 ]$ & $0.44 \pm 0.02$ & $0.40 \pm 0.11$   & $0.81 \pm 0.16$    & $36.1 \pm 5.5$      & $1.4 \pm 0.4$   & $11\pm 1$           &$0.16 \pm 0.06$\\
$[ 43 , 50 ]$ & $< 0.05$             & $<0.3$                  & $<0.3$                    & $2.7 \pm 0.1$   & $< 1.0$           &  $0.96 \pm 0.2$   &$<0.15$\\
\hline
G5.89-0.4 \\
$[15 , 25]$   &  $0.15 \pm 0.02$ & $0.96 \pm 0.13$ & $0.45 \pm 0.1$      & $12.7 \pm 0.5$  & $1.3 \pm 0.2$ & $5 \pm 0.9$           &$< 0.6$  \\
\hline
B1741-312 \\
$[-7 , 7]$ & $0.23 \pm 0.05$ &  $0.81 \pm 0.4$      & $<0.5$                      & $23\pm 0.5 $ & $...$                   & $5.5 \pm 0.2$            & $...$\\
\hline
B1730-130 \\
$[0 , 10] $& $0.06 \pm 0.02$ & $0.06\pm 0.03$ & $<0.2$                           & $1.2\pm 0.1 $ & $...$                 & $0.05\pm 0.02$  &$...$\\
\hline
\end{tabular}
\tablefoot{The first column gives the velocity intervals used for deriving the various molecular column densities listed in the other columns.
}
\end{table*}

\begin{table*}
\label{tab:abundances}
\caption{Abundances relative to H$_2$ }
\begin{tabular}{lcclc}
\hline
Molecule & Abundance & Uncertainty & Comment & $N($H$_2$$)$/$\int \tau dv$ \\
                &                    &  dex         &                    &  cm$^{-2}$/\kms \\
\hline
HCO$^+$ & $3.1 \times 10^{-9}$ 	& 0.21 	&  using [CH] = $3.6 \times 10^{-8}$ and E(B-V)$^e$ & $4.0 \times 10^{20}$ \\
HOC$^+$ & $4.6 \times 10^{-11}$ 	& 0.21 	& from [HOC$^+$]/[HCO$^+$] 				& $5.2 \times 10^{22}$\\
CF$^+$ & $1.7 \times 10^{-10}$ 	& 0.30 	& from [CF$^+$]/[HCO$^+$] 					& $9.0 \times 10^{22}$\\
C$_3$H$^+$ & $7.5 \times 10^{-11}$ & 0.30 	& from [C$_3$H$^+$]/[HCO$^+$]				& $2.0 \times 10^{23}$\\
\hline
HF & $1.2 \times 10^{-8}$ & 0.14 & using [CH] = $3.6 \times 10^{-8}$$^e$					& $2.0 \times 10^{20}$\\
H$_2$O$^a$ & $2.7 \times 10^{-8}$ & 0.20 & using [CH] = $3.6 \times 10^{-8}$$^e$ 				& $3.4 \times 10^{20}$\\
CCH & $4.4 \times 10^{-8}$ & 0.15 & using [CH] = $3.6 \times 10^{-8}$$^e$ 					& $1.5 \times 10^{21}$\\
CH$^b$   & $3.6 \times 10^{-8}$  &    0.21     &    \cite{sheffer:08}                                                         &     $1.0 \times 10^{21}$\\ 
CH$^c$   & $3.6 \times 10^{-8}$  &  0.21       &    \cite{sheffer:08}                                                         &     $9.7 \times 10^{20}$\\ 
OH$^d$ &  $1.0 \times 10^{-7}$&    0.1        & \cite{weselak:10}  &                                $2.5\times 10^{20}$                                  \\                          
\hline
\end{tabular}
\tablefoot{$^a$ The entry in the  last column refers to the ground state p-H$_2$O line at 1.13~THz and assume an ortho-to-para ratio of three.
 $^b$ The entry in the last column refers to the 532 or the 536~GHz transitions of CH.
 $^c$ The entry in the last column refers to the 2~THz  transitions of CH.
 $^d$ The entry in the last column refers to the 2.5~THz  transitions of OH. \cite{weselak:10} present data for five objects only.
$^e$ \cite{sheffer:08}.
}
\end{table*}

\section {Chemistry}
\label{sec:chemistry}
The revised determinations of the molecular ion abundances, and the tight relationships between species provide
new constraints on the formation and destruction pathways of the studied molecular ions.
As discussed by \citet{liszt:04}, HOC$^+$ in diffuse gas is mainly produced by the exothermic reaction between C$^+$ and H$_2$O.
This reaction also produces HCO$^+$ but  is thought to always be a minor formation pathway for this ion. A second production
route for both HOC$^+$ and HCO$^+$ is the reaction between CO$^+$ and H$_2$, CO$^+$ itself being mostly formed in the reaction between C$^+$ and OH.
In diffuse cloud conditions, HCO$^+$ and HOC$^+$ are both destroyed by dissociative
recombination  with electrons,  HOC$^+$ is also destroyed in the isomerization reaction with H$_2$ which produces HCO$^+$, the two processes having similar rates.
Following \cite{liszt:04}, and using the chemical scheme described in Appendix \ref{app:chem}, the ratios of the HOC$^+$ and H$_2$O abundances 
at steady state can be written :

\begin{equation}
\frac{[HOC^+]}{[H_2O]} = x_C \frac{2 k_1(k'_e x_C + k_3 f(H_2))  + k_3 k_2 f(H_2) [OH]/[H_2O]}{(k'_e x_C  + k_3 f(H_2))(2k_e x_C + k_4f(H_2))}
\end{equation}

The behavior of this ratio with temperature and H$_2$ fraction is illustrated in Figures \ref{fig:rates}, \ref{fig:rates1} and \ref{fig:rates2} in which we used the observed [OH]/[H$_2$O] ratio ($\sim 4$)  quoted
above, and two hypotheses for the rate coefficients, the Langevin rate or an enhanced capture rate at low temperature for ion molecule reactions \citep{woon:09}.
 The tight relationship between HOC$^+$, OH and H$_2$O and the observed abundance ratio are qualitatively consistent
with this simple chemical scheme. We note that the observed value of $N($HOC$^+$$))/N($H$_2$O$) \sim 2 \times 10^{-3}$ is in better agreement with the lower values of the rate
coefficients, which do not include the enhancement of the capture rate in ion-polar molecule reactions. 
Because of the important role reactions (R1) between C$^+$ and H$_2$O, and (R2) between C$^+$ and OH, play in initiating the  diffuse gas chemistry,
a better knowledge of their rate coefficients, their temperature dependence and branching ratios would be very beneficial.

It is not possible to understand the observed relationship between HCO$^+$ and H$_2$O with this simple chemical scheme because it
predicts a significantly lower [HCO$^+$]/[H$_2$O] abundance ratio than observed, in agreement with the existence of other, 
more dominant, chemical pathways for HCO$^+$ that are initiated by CH$^+$ and do not lead to HOC$^+$ \citep{godard:09,godard:14}. 
Whatever the chemical pathways that are involved, they must predict the tight relationship between
HCO$^+$ and OH or H$_2$O. For instance \citet{godard:09} have shown that predicted  HCO$^+$ and OH abundances  by the turbulent dissipation regions (TDR) 
models satisfactorily reproduce the  observed trend.

The last panel of Fig \ref{fig:rates1} and \ref{fig:rates2}  presents the expected abundance ratio [CF$^+$]/[HF]. 
The presence of detectable quantities of CF$^+$  was predicted by \citet{neufeld:05}, before its actual detection in
the Orion Bar \citep{neufeld:06}. As discussed previously \citep{nw:09,liszt:14,guzman:12}, the
CF$^+$ chemistry is very simple and, in diffuse gas where the electrons are produced by the ionization of carbon, the
  abundance ratio  [CF$^+$]/[HF] can be simply written as
 
  \begin{equation}
 [CF^+]/[HF] = k_5/kr_e = 0.053 (T/70~K)^{0.65} 
\end{equation}

using the reaction rates from \citet{nw:09} as listed in Table \ref{tab:rates}. Using instead the Langevin rate for the reaction R5 (C$^+$ + HF )
leads to a very low value for [CF$^+$]/[HF] of 0.5\% at 70~K (see Fig \ref{fig:rates1}). Using  the enhanced capture rate
for ion-molecule reactions leads to  a rate coefficient of $\sim 5 \times 10^{-9}$ cm$^3$s$^{-1}$ at 100~K  increasing with decreasing 
temperatures, and a predicted value of  [CF$^+$]/[HF]  of 3.4 \% at 70~K (see Fig \ref{fig:rates}), in better agreement with the observations but
still somewhat too high. A recent theoretical study has been presented by \citet{denis-alpizar:18} for the temperature range 50 - 2000~K.  They predict a rate coefficient only moderately larger than the Langevin rate, which reaches
$1.35 \times 10^{-9}$ cm$^3$s$^{-1}$ at 70~K and $1.22 \times 10^{-9}$ cm$^3$s$^{-1}$ at 100~K (see Table \ref{tab:rates}). The corresponding [CF$^+$]/[HF] ratio is 0.8\%, a factor of two lower than the mean observed value, but 
still marginally consistent given  the large uncertainties. Therefore the data and the recent theoretical calculations agree that the rate coefficient of reaction R5 used by \citet{nw:09} is too high and 
should be revised downward by a factor of about $\sim 5.3/1.7 \sim 3$.  
A further test of the chemistry would be to search for variations of the abundance ratio with the temperature, but the current
data quality are too  marginal at this stage. The W49N sightline is a promising target because of the relative spread in kinetic
temperatures along the line of sight, with a cold and narrow  velocity component near 40 kms$^{-1}$ , prominent in CF$^+$ \citep{gerin:15,liszt:15}.  

\begin{table}
\caption{Summary of abundance ratios}
\label{tab:ratio}
\begin{tabular}{lcc}
\hline
Ratio & Galactic Disk &   SgrB2   \\
          &                       & $V_{LSR} < -20$\, \kms \\
\hline 
HF/p-H$_2$O                     & $1.9                            \pm 0.4 $                       &            $  0.43             \pm                 0.1$ \\
HCO$^+$/p-H$_2$O       & $0.47                          \pm 0.03   $                       &             $ 0.37           \pm                 0.03 $\\
HCO$^+$/HF                  & $0.25 \pm                     0.02$                                &     $0.88                     \pm 0.1$                \\ 
HOC$^+$/HCO$^+$      & $(1.5 \pm 0.2) \times 10^{-2}$  & $( 0.8 \pm 0.1) \times 10^{-2} $ \\ 
CF$^+$/HF                    & $(1.7 \pm 0.3) \times 10^{-2}$  & $(3.0 \pm 0.3) \times 10^{-2} $  \\
\hline
\end{tabular}
\end{table}

\section{Material observed toward SgrA and SgrB2}

Although sight-lines toward SgrA (G0.02-0.07) and SgrB2 were included in the observing program (Table \ref{tab:sources}), 
abundances in the material sampled along these directions were not included 
when deriving the relationships and abundances discussed in Sect. 3 and shown in 
Table \ref{tab:abundances}.  Gas seen in absorption at negative velocities toward SgrA and SgrB2 arises in the 
central molecular zone within about 200 pc of the Galactic center and exists under conditions of 
different (mostly higher) density, pressure, temperature, metallicity and cosmic-ray 
ionization rate \citep{morris:96,goto:14}. 
It is unclear whether any of the material observed along these lines of sight can be 
considered as diffuse molecular gas in the usual sense that is applicable to
Galactic disk gas, that is being optically transparent and 
having moderate density and pressure \citep{thiel:17,liszt:18a}.

Absolute abundances cannot be accurately determined because the abundances of the usual 
H$_2$-tracers (CH, HF, HCO$^+$) with respect to H$_2$ have not been accurately established 
in the Galactic Center region. \citet{sonnentrucker:13} have shown that H$_2$O and CH are enhanced relative to HF along sight-lines close to SgrA*, but not by the
same factor ($\sim 5$ for H$_2$O and $\sim 2$ for CH). It is therefore expected that the  various molecular abundance ratios displayed in Table 5
reflect the differences in the physical conditions between the Galactic disk and the central molecular zone.
The N(HOC$^+$)/N(HCO$^+$) and 
N(HOC$^+$)/N(p-H$_2$O) ratios near the Galactic nucleus are 55\% of what is seen 
in the disk, while the N(HCO$^+$)/N(p-H$_2$O) ratio also declines, but only by 30\%.  
More extreme, \cite{sonnentrucker:15} found that N(HF)/N(H$_2$O) is a factor 
4.4 smaller in the Galactic center and our data show a comparable decline by a 
factor 3.5 in the ratio N(HF)/N(HCO$^+$).  By contrast, CF$^+$
seems to vary somewhat less between the disk and center: N(CF$^+$)/N(HOC$^+$)
is larger in the Galactic center by about the same factor by which N(HOC$^+$)/N(HCO$^+$)
declines, and N(CF$^+$)/N(HF) increases by 75\%, consistent with the decline in
N(HF)/N(H$_2$O).

All things considered, the biggest surprise here may be the relative insensitivity of 
the chemistry to the strong differences in conditions that are supposed to occur near 
the Galactic center.  According to the discussion in Sect. 5 (see Fig. C.1) a lower
N(HOC$^+$)/N(p-H$_2$O) ratio can be explained with a  higher molecular fraction in the 
gas but the higher kinetic temperatures expected in the central molecular zone would
have an opposite effect.

\section{Summary and conclusions}
\label{sec:conclusion}

Using ALMA Cycle 2 data we discussed observations of the molecular ions CF$^+$, HCO$^+$, HOC$^+$ and C$_3$H$^+$ in 
absorption along inner-galaxy sight-lines toward a variety of background sources listed in Table 1: the 
distant quasar B1730-130 aka J1733-1304, two compact HII regions G5.89-0.4 and G10.62-0.38 in the galactic disk, 
and the HII regions SgrA G0.02-0.07, SgrB2-M and SgrB2-S in the central molecular zone. 
Neutral species HCN, HNC, $c$-C$_3$H, NH$_2$CHO and $^{13}$CS were also observed in the same program 
(Table 2, Fig. B.1) but were not discussed in any detail. Observations of absorption toward another 
quasar in the same observing program, B1741-312, were presented in a previous paper that
discussed the molecular contribution from gas in the Galactic bulge, outside the central
molecular zone  \citep{gerin:17}.

Spectra of the species we observed are shown in Fig. 1 and Fig. B.1.  Fig. 2 compares spectra of
the ions HCO$^+$ and HOC$^+$ with those of HF and H$_2$O toward W31 G10.6-0.38, highlighting
the strong similarity that was observed among the oxygen-bearing species in this work.  Fig. 3
shows spectral channel-by-channel graphs of the interrelationships among these species and with
CF$^+$ whose abundance is proportional to that of HF.  For future reference, Table 3 gives
column densities for all observed species over selected velocity ranges toward all sources 
but the abundances of the neutral species observed here were not discussed in this paper.

Absorption originating in the central molecular zone and seen at negative velocities toward SgrA 
and SgrB2 has abundance patterns that differ from those of the Galactic disk gas,  and was not 
included in the average  molecular abundances and abundance ratios discussed in Sect. 3 and 
shown in Table 4.  The Galactic disk and Galactic center abundances are summarized separately.  

\subsection{Gas in the Galactic disk}

Using the known abundance of N(CH)/N(H$_2$) $= 3.6\times10^{-8}$ and Herschel HiFi measurements of  ground state transitions of CH at 532 and 536~GHz,
we compared HCO$^+$ and CH absorption profiles to derive the relative abundance N(HCO$^+$)/N(H$_2$) 
$= (3.1\pm1.9) \times10^{-9}$,  confirming the previously-determined ratio  $3 \times10^{-9}$.  We used this 
to determine the abundances of the other ions CF$^+$, HOC$^+$, and C$_3$H$^+$ relative to H$_2$ as shown 
in Table 4. Table 4 also gives the relative abundances of HF, H$_2$O, CCH, CH, and OH that were used as
comparisons in discussion of the molecular chemistry.

C$_3$H$^+$ was found to be common in our sight-line sample in diffuse molecular gas in the Galactic disk, the first time this
species has been seen outside the Orion A and Orion B photo-dissociation regions and the central molecular zone.  The relative 
abundance N(C$_3$H$^+$)/N(H$_2$) $= 7.5 \times 10^{-11}$ is similar to that seen in Orion. Absorption by 
C$_3$H$^+$ is weak in the transition observed here and is contaminated by the spectra of other species especially
toward SgrB2. This limits the usefulness of C$_3$H$^+$ as a general tracer but the constancy of its 
observed abundance should be further checked, and could provide a useful constraint on chemical modeling.

We also confirmed a lower CF$^+$ abundance N(CF$^+$)/N(H$_2$) $= 1.5 \times 10^{-10}$ than was
originally predicted, and we ascribed it to a slow formation rate of the reaction C$^+$ + 
HF $\rightarrow$ CF$^+$ + H (Sect. 5) as recently calculated by \citet{denis-alpizar:18}.  Although CF$^+$ has a near-constant abundance with respect 
to the H$_2$-tracer HF, absorption in the transition that we observed (Table 2) is weak and 
contaminated by an overlapping hydrogen recombination line toward HII regions, 
compromising its usefulness to trace H$_2$.

The abundances and profiles of HOC$^+$, HCO$^+$ and H$_2$O are tightly correlated in the Galactic
disk gas (see Fig. 2), with N(HOC$^+$)/N(HCO$^+$) $= (1.5\pm 0.3) \times 10^{-2}$ and 
N(HCO$^+$)/N(p-H$_2$O) $= 0.47\pm0.03$.  The tight association of HCO$^+$ with the other two species
is puzzling in that the abundance of HCO$^+$ in the diffuse gas requires a turbulent-dissipative formation chemistry 
\citep{godard:14} and the channel that forms HCO$^+$ in the reaction of C$^+$ and H$_2$O is not an important
source of HCO$^+$, while the HOC$^+$/H$_2$O ratio is best explained by the reaction of C$^+$ and  H$_2$O at Langevin rates 
(Sect. 5 and Fig. C.1) and the column densities of water are also explained by the quiescent
chemistry.  The rate coefficients of the reactions between C$^+$ and OH and H$_2$O have an important effect on the predicted
chemical abundances and their absolute value and temperature dependence should be more accurately known.
   
\subsection{Gas observed toward SgrB2}

Gas seen in absorption at negative velocities toward SgrA and  SgrB2 arises in the central
molecular zone within 200 pc of the Galactic center and exists under conditions of 
different (mostly higher) density, pressure, temperature, metallicity and cosmic-ray 
ionization rate as discussed in Sect. 6. This is presumably reflected in the molecular 
abundance ratios that differ from those in the galactic disk by as much as a factor two
for the N(HOC$^+$)/N(HCO$^+$) ratio or 4 for N(HF)/N(H$_2$O) and N(HF)/N(HCO$^+$).  Absolute 
abundances cannot be determined in the central molecular 
zone because the relative abundances of the usual H$_2$-tracers (CH, HF, etc) have not been 
established in this region, and it is unclear whether any of the gas in the
central molecular zone is diffuse in the sense that is usually understood in the Galactic disk,
being optically transparent and having moderate density and pressure. The spectra of CF$^+$
and C$_3$H$^+$ are contaminated along the galactic center sight-lines where the optical depths of
abundant species like HF and HCO$^+$ are high, complicating the derivation of accurate
molecular abundances.


\begin{acknowledgements}
This paper makes use of the following ALMA data: ADS/JAO.ALMA\#2013.1.01194.S. ALMA is a partnership of ESO (representing its member states), NSF (USA) and NINS (Japan), together with NRC (Canada), NSC and ASIAA (Taiwan), and KASI (Republic of Korea), in cooperation with the Republic of Chile. The Joint ALMA Observatory is operated by ESO, AUI/NRAO and NAOJ. The National Radio Astronomy Observatory is a facility of the
National Science Foundation operated under cooperative agreement by Associated Universities, Inc. We acknowledge funding support from 
PCMI-CNRS/INSU and AS ALMA from Observatoire de Paris. We thank E. Bergin and E. Falgarone for illuminating discussions, and the referee for a careful reading of the manuscript.
  \end{acknowledgements}

 \bibliographystyle{aa}
\bibliography{alma-ion}

\begin{thebibliography}{60}
\expandafter\ifx\csname natexlab\endcsname\relax\def\natexlab#1{#1}\fi

\bibitem[{{Cuadrado} {et~al.}(2015){Cuadrado}, {Goicoechea}, {Pilleri},
  {Cernicharo}, {Fuente}, \& {Joblin}}]{cuadrado:15}
{Cuadrado}, S., {Goicoechea}, J.~R., {Pilleri}, P., {et~al.} 2015, \aap, 575,
  A82

\bibitem[{{Denis-Alpizar} {et~al.}(2018){Denis-Alpizar}, {Guzm{\'a}n}, \&
  {Inostroza}}]{denis-alpizar:18}
{Denis-Alpizar}, O., {Guzm{\'a}n}, V.~V., \& {Inostroza}, N. 2018, \mnras, 479,
  753

\bibitem[{{Flagey} {et~al.}(2013){Flagey}, {Goldsmith}, {Lis}, {Gerin},
  {Neufeld}, {Sonnentrucker}, {De Luca}, {Godard}, {Goicoechea}, {Monje}, \&
  {Phillips}}]{flagey:13}
{Flagey}, N., {Goldsmith}, P.~F., {Lis}, D.~C., {et~al.} 2013, \apj, 762, 11

\bibitem[{{Gerin} {et~al.}(2010){Gerin}, {de Luca}, {Goicoechea}, {Herbst},
  {Falgarone}, {Godard}, {Bell}, {Coutens}, {Ka{\'z}mierczak}, {Sonnentrucker},
  {Black}, {Neufeld}, {Phillips}, {Pearson}, {Rimmer}, {Hassel}, {Lis},
  {Vastel}, {Boulanger}, {Cernicharo}, {Dartois}, {Encrenaz}, {Giesen},
  {Goldsmith}, {Gupta}, {Gry}, {Hennebelle}, {Hily-Blant}, {Joblin},
  {Ko{\l}os}, {Kre{\l}owski}, {Mart{\'{\i}}n-Pintado}, {Monje}, {Mookerjea},
  {Perault}, {Persson}, {Plume}, {Salez}, {Schmidt}, {Stutzki}, {Teyssier},
  {Yu}, {Contursi}, {Menten}, {Geballe}, {Schlemmer}, {Morris}, {Hatch},
  {Imram}, {Ward}, {Caux}, {G{\"u}sten}, {Klein}, {Roelfsema}, {Dieleman},
  {Schieder}, {Honingh}, \& {Zmuidzinas}}]{gerin:10}
{Gerin}, M., {de Luca}, M., {Goicoechea}, J.~R., {et~al.} 2010, \aap, 521, L16

\bibitem[{{Gerin} {et~al.}(2011){Gerin}, {Ka{\'z}mierczak}, {Jastrzebska},
  {Falgarone}, {Hily-Blant}, {Godard}, \& {de Luca}}]{gerin:11}
{Gerin}, M., {Ka{\'z}mierczak}, M., {Jastrzebska}, M., {et~al.} 2011, \aap,
  525, A116

\bibitem[{{Gerin} \& {Liszt}(2017)}]{gerin:17}
{Gerin}, M. \& {Liszt}, H. 2017, \aap, 600, A48

\bibitem[{{Gerin} {et~al.}(2016){Gerin}, {Neufeld}, \& {Goicoechea}}]{gerin:16}
{Gerin}, M., {Neufeld}, D.~A., \& {Goicoechea}, J.~R. 2016, \araa, 54, 181

\bibitem[{{Gerin} {et~al.}(2015){Gerin}, {Ruaud}, {Goicoechea}, {Gusdorf},
  {Godard}, {de Luca}, {Falgarone}, {Goldsmith}, {Lis}, {Menten}, {Neufeld},
  {Phillips}, \& {Liszt}}]{gerin:15}
{Gerin}, M., {Ruaud}, M., {Goicoechea}, J.~R., {et~al.} 2015, \aap, 573, A30

\bibitem[{{Godard} {et~al.}(2010){Godard}, {Falgarone}, {Gerin}, {Hily-Blant},
  \& {de Luca}}]{godard:10}
{Godard}, B., {Falgarone}, E., {Gerin}, M., {Hily-Blant}, P., \& {de Luca}, M.
  2010, \aap, 520, A20

\bibitem[{{Godard} {et~al.}(2012){Godard}, {Falgarone}, {Gerin}, {Lis}, {De
  Luca}, {Black}, {Goicoechea}, {Cernicharo}, {Neufeld}, {Menten}, \&
  {Emprechtinger}}]{godard:12}
{Godard}, B., {Falgarone}, E., {Gerin}, M., {et~al.} 2012, \aap, 540, A87

\bibitem[{{Godard} {et~al.}(2009){Godard}, {Falgarone}, \& {Pineau Des
  For{\^e}ts}}]{godard:09}
{Godard}, B., {Falgarone}, E., \& {Pineau Des For{\^e}ts}, G. 2009, \aap, 495,
  847

\bibitem[{{Godard} {et~al.}(2014){Godard}, {Falgarone}, \& {Pineau des
  For{\^e}ts}}]{godard:14}
{Godard}, B., {Falgarone}, E., \& {Pineau des For{\^e}ts}, G. 2014, \aap, 570,
  A27

\bibitem[{{Goss}(1968)}]{goss:68}
{Goss}, W.~M. 1968, \apjs, 15, 131

\bibitem[{{Goto} {et~al.}(2014){Goto}, {Geballe}, {Indriolo}, {Yusef-Zadeh},
  {Usuda}, {Henning}, \& {Oka}}]{goto:14}
{Goto}, M., {Geballe}, T.~R., {Indriolo}, N., {et~al.} 2014, \apj, 786, 96

\bibitem[{{Green} {et~al.}(2015){Green}, {Schlafly}, {Finkbeiner}, {Rix},
  {Martin}, {Burgett}, {Draper}, {Flewelling}, {Hodapp}, {Kaiser}, {Kudritzki},
  {Magnier}, {Metcalfe}, {Price}, {Tonry}, \& {Wainscoat}}]{green:15}
{Green}, G.~M., {Schlafly}, E.~F., {Finkbeiner}, D.~P., {et~al.} 2015, \apj,
  810, 25

\bibitem[{{Guzm{\'a}n} {et~al.}(2012){Guzm{\'a}n}, {Pety}, {Gratier},
  {Goicoechea}, {Gerin}, {Roueff}, \& {Teyssier}}]{guzman:12}
{Guzm{\'a}n}, V., {Pety}, J., {Gratier}, P., {et~al.} 2012, \aap, 543, L1

\bibitem[{{Guzm{\'a}n} {et~al.}(2015){Guzm{\'a}n}, {Pety}, {Goicoechea},
  {Gerin}, {Roueff}, {Gratier}, \& {{\"O}berg}}]{guzman:15}
{Guzm{\'a}n}, V.~V., {Pety}, J., {Goicoechea}, J.~R., {et~al.} 2015, \apjl,
  800, L33

\bibitem[{{Hamberg} {et~al.}(2014){Hamberg}, {Kashperkat}, {Thomas}, {Roueff},
  {Zhaunerchyk}, {Danielsson}, {af Ugglast}, {\"Osterdahl}, {Vigren},
  {Kaminska}, {K\"allberg}, {Siminsson}, {Paal}, {Gerin}, {Larsson}, \&
  {Geppert}}]{hamberg:14}
{Hamberg}, M., {Kashperkat}, I., {Thomas}, R.~D., {et~al.} 2014, J. Phys. Chem.
  A., 118, 6034

\bibitem[{{Indriolo} {et~al.}(2013){Indriolo}, {Neufeld}, {Seifahrt}, \&
  {Richter}}]{indriolo:13}
{Indriolo}, N., {Neufeld}, D.~A., {Seifahrt}, A., \& {Richter}, M.~J. 2013,
  \apj, 764, 188

\bibitem[{{Jenkins} \& {Tripp}(2011)}]{jenkins:11}
{Jenkins}, E.~B. \& {Tripp}, T.~M. 2011, \apj, 734, 65

\bibitem[{{Lallement} {et~al.}(2014){Lallement}, {Vergely}, {Valette},
  {Puspitarini}, {Eyer}, \& {Casagrande}}]{lallement:14}
{Lallement}, R., {Vergely}, J.-L., {Valette}, B., {et~al.} 2014, \aap, 561, A91

\bibitem[{{Liszt} {et~al.}(2018){Liszt}, {Gerin}, {Beasley}, \&
  {Pety}}]{liszt:18a}
{Liszt}, H., {Gerin}, M., {Beasley}, A., \& {Pety}, J. 2018, \apj, 856, 151

\bibitem[{{Liszt} {et~al.}(2004){Liszt}, {Lucas}, \& {Black}}]{liszt:04}
{Liszt}, H., {Lucas}, R., \& {Black}, J.~H. 2004, \aap, 428, 117

\bibitem[{{Liszt} \& {Gerin}(2016)}]{liszt:16}
{Liszt}, H.~S. \& {Gerin}, M. 2016, \aap, 585, A80

\bibitem[{{Liszt} {et~al.}(2015){Liszt}, {Guzm{\'a}n}, {Pety}, {Gerin},
  {Neufeld}, \& {Gratier}}]{liszt:15}
{Liszt}, H.~S., {Guzm{\'a}n}, V.~V., {Pety}, J., {et~al.} 2015, \aap, 579, A12

\bibitem[{{Liszt} {et~al.}(2014){Liszt}, {Pety}, {Gerin}, \&
  {Lucas}}]{liszt:14}
{Liszt}, H.~S., {Pety}, J., {Gerin}, M., \& {Lucas}, R. 2014, \aap, 564, A64

\bibitem[{{Lucas} \& {Liszt}(1996)}]{lucas:96}
{Lucas}, R. \& {Liszt}, H. 1996, \aap, 307, 237

\bibitem[{{Martinez} {et~al.}(2008){Martinez}, {Betts}, {Villano}, {Eyet},
  {Snow}, \& {Bierbaum}}]{martinez:08}
{Martinez}, Jr., O., {Betts}, N.~B., {Villano}, S.~M., {et~al.} 2008, \apj,
  686, 1486

\bibitem[{{McGuire} {et~al.}(2013){McGuire}, {Carroll}, {Loomis}, {Blake},
  {Hollis}, {Lovas}, {Jewell}, \& {Remijan}}]{mcguire:13}
{McGuire}, B.~A., {Carroll}, P.~B., {Loomis}, R.~A., {et~al.} 2013, \apj, 774,
  56

\bibitem[{{Morris} \& {Serabyn}(1996)}]{morris:96}
{Morris}, M. \& {Serabyn}, E. 1996, \araa, 34, 645

\bibitem[{{M{\"u}ller} {et~al.}(2016){M{\"u}ller}, {Belloche}, {Xu}, {Lees},
  {Garrod}, {Walters}, {van Wijngaarden}, {Lewen}, {Schlemmer}, \&
  {Menten}}]{muller:16}
{M{\"u}ller}, H.~S.~P., {Belloche}, A., {Xu}, L.-H., {et~al.} 2016, \aap, 587,
  A92

\bibitem[{{Neufeld} {et~al.}(2015){Neufeld}, {Godard}, {Gerin}, {Pineau des
  For{\^e}ts}, {Bernier}, {Falgarone}, {Graf}, {G{\"u}sten}, {Herbst},
  {Lesaffre}, {Schilke}, {Sonnentrucker}, \& {Wiesemeyer}}]{neufeld:15}
{Neufeld}, D.~A., {Godard}, B., {Gerin}, M., {et~al.} 2015, \aap, 577, A49

\bibitem[{{Neufeld} {et~al.}(2002){Neufeld}, {Kaufman}, {Goldsmith},
  {Hollenbach}, \& {Plume}}]{neufeld:02}
{Neufeld}, D.~A., {Kaufman}, M.~J., {Goldsmith}, P.~F., {Hollenbach}, D.~J., \&
  {Plume}, R. 2002, \apj, 580, 278

\bibitem[{{Neufeld} {et~al.}(2006){Neufeld}, {Schilke}, {Menten}, {Wolfire},
  {Black}, {Schuller}, {M{\"u}ller}, {Thorwirth}, {G{\"u}sten}, \&
  {Philipp}}]{neufeld:06}
{Neufeld}, D.~A., {Schilke}, P., {Menten}, K.~M., {et~al.} 2006, \aap, 454, L37

\bibitem[{{Neufeld} \& {Wolfire}(2009)}]{nw:09}
{Neufeld}, D.~A. \& {Wolfire}, M.~G. 2009, \apj, 706, 1594

\bibitem[{{Neufeld} {et~al.}(2005){Neufeld}, {Wolfire}, \&
  {Schilke}}]{neufeld:05}
{Neufeld}, D.~A., {Wolfire}, M.~G., \& {Schilke}, P. 2005, \apj, 628, 260

\bibitem[{{Novotny} {et~al.}(2005){Novotny}, {Mitchell}, {LeGarrec},
  {Florescu-Mitchell}, {Rebrion-Rowe}, {Svendsen}, {El Ghazaly}, {Andersen},
  {Ehlerding}, {Viggiano}, {Hellberg}, {Thomas}, {Zhaunerchyk}, {Geppert},
  {Montaigne}, {Kaminska}, {{\"O}sterdahl}, \& {Larsson}}]{novotny:05}
{Novotny}, O., {Mitchell}, J.~B.~A., {LeGarrec}, J.~L., {et~al.} 2005, Journal
  of Physics B Atomic Molecular Physics, 38, 1471

\bibitem[{{Pearson} {et~al.}(2008){Pearson}, {Brauer}, \&
  {Drouin}}]{Pearson:08}
{Pearson}, J.~C., {Brauer}, C.~S., \& {Drouin}, B.~J. 2008, Journal of
  Molecular Spectroscopy, 251, 394

\bibitem[{{Pety} {et~al.}(2012){Pety}, {Gratier}, {Guzm{\'a}n}, {Roueff},
  {Gerin}, {Goicoechea}, {Bardeau}, {Sievers}, {Le Petit}, {Le Bourlot},
  {Belloche}, \& {Talbi}}]{pety:12}
{Pety}, J., {Gratier}, P., {Guzm{\'a}n}, V., {et~al.} 2012, \aap, 548, A68

\bibitem[{{Pineda} {et~al.}(2013){Pineda}, {Langer}, {Velusamy}, \&
  {Goldsmith}}]{pineda:13}
{Pineda}, J.~L., {Langer}, W.~D., {Velusamy}, T., \& {Goldsmith}, P.~F. 2013,
  \aap, 554, A103

\bibitem[{{Planck Collaboration} {et~al.}(2011){Planck Collaboration},
  {Abergel}, {Ade}, {Aghanim}, {Arnaud}, {Ashdown}, {Aumont}, {Baccigalupi},
  {Balbi}, {Banday}, \& et~al.}]{abergel:11}
{Planck Collaboration}, {Abergel}, A., {Ade}, P.~A.~R., {et~al.} 2011, \aap,
  536, A24

\bibitem[{{Plume} {et~al.}(2004){Plume}, {Kaufman}, {Neufeld}, {Snell},
  {Hollenbach}, {Goldsmith}, {Howe}, {Bergin}, {Melnick}, \&
  {Bensch}}]{plume:04}
{Plume}, R., {Kaufman}, M.~J., {Neufeld}, D.~A., {et~al.} 2004, \apj, 605, 247

\bibitem[{{Rachford} {et~al.}(2009){Rachford}, {Snow}, {Destree}, {Ross},
  {Ferlet}, {Friedman}, {Gry}, {Jenkins}, {Morton}, {Savage}, {Shull},
  {Sonnentrucker}, {Tumlinson}, {Vidal-Madjar}, {Welty}, \&
  {York}}]{rachford:09}
{Rachford}, B.~L., {Snow}, T.~P., {Destree}, J.~D., {et~al.} 2009, \apjs, 180,
  125

\bibitem[{{Remy} {et~al.}(2017){Remy}, {Grenier}, {Marshall}, \&
  {Casandjian}}]{remy:17}
{Remy}, Q., {Grenier}, I.~A., {Marshall}, D.~J., \& {Casandjian}, J.~M. 2017,
  \aap, 601, A78

\bibitem[{{Schlegel} {et~al.}(1998){Schlegel}, {Finkbeiner}, \&
  {Davis}}]{schlegel:98}
{Schlegel}, D.~J., {Finkbeiner}, D.~P., \& {Davis}, M. 1998, \apj, 500, 525

\bibitem[{{Sheffer} {et~al.}(2008){Sheffer}, {Rogers}, {Federman}, {Abel},
  {Gredel}, {Lambert}, \& {Shaw}}]{sheffer:08}
{Sheffer}, Y., {Rogers}, M., {Federman}, S.~R., {et~al.} 2008, \apj, 687, 1075

\bibitem[{{Smith} {et~al.}(2002){Smith}, {Schlemmer}, {von Richthofen}, \&
  {Gerlich}}]{smith:02}
{Smith}, M.~A., {Schlemmer}, S., {von Richthofen}, J., \& {Gerlich}, D. 2002,
  \apjl, 578, L87

\bibitem[{{Snow} \& {McCall}(2006)}]{snow:06}
{Snow}, T.~P. \& {McCall}, B.~J. 2006, \araa, 44, 367

\bibitem[{{Sonnentrucker} {et~al.}(2013){Sonnentrucker}, {Neufeld}, {Gerin},
  {De Luca}, {Indriolo}, {Lis}, \& {Goicoechea}}]{sonnentrucker:13}
{Sonnentrucker}, P., {Neufeld}, D.~A., {Gerin}, M., {et~al.} 2013, \apjl, 763,
  L19

\bibitem[{{Sonnentrucker} {et~al.}(2015){Sonnentrucker}, {Wolfire}, {Neufeld},
  {Flagey}, {Gerin}, {Goldsmith}, {Lis}, \& {Monje}}]{sonnentrucker:15}
{Sonnentrucker}, P., {Wolfire}, M., {Neufeld}, D.~A., {et~al.} 2015, \apj, 806,
  49

\bibitem[{{Spitzer} \& {Jenkins}(1975)}]{spitzer:75}
{Spitzer}, Jr., L. \& {Jenkins}, E.~B. 1975, \araa, 13, 133

\bibitem[{{Stoffels} {et~al.}(2016){Stoffels}, {Kluge}, {Schlemmer}, \&
  {Br{\"u}nken}}]{stoffels:16}
{Stoffels}, A., {Kluge}, L., {Schlemmer}, S., \& {Br{\"u}nken}, S. 2016, \aap,
  593, A56

\bibitem[{{Thiel} {et~al.}(2017){Thiel}, {Belloche}, {Menten}, {Garrod}, \&
  {M{\"u}ller}}]{thiel:17}
{Thiel}, V., {Belloche}, A., {Menten}, K.~M., {Garrod}, R.~T., \& {M{\"u}ller},
  H.~S.~P. 2017, \aap, 605, L6

\bibitem[{{Wakelam} {et~al.}(2012){Wakelam}, {Herbst}, {Loison}, {Smith},
  {Chandrasekaran}, {Pavone}, {Adams}, {Bacchus-Montabonel}, {Bergeat},
  {B{\'e}roff}, {Bierbaum}, {Chabot}, {Dalgarno}, {van Dishoeck}, {Faure},
  {Geppert}, {Gerlich}, {Galli}, {H{\'e}brard}, {Hersant}, {Hickson},
  {Honvault}, {Klippenstein}, {Le Picard}, {Nyman}, {Pernot}, {Schlemmer},
  {Selsis}, {Sims}, {Talbi}, {Tennyson}, {Troe}, {Wester}, \&
  {Wiesenfeld}}]{wakelam:12}
{Wakelam}, V., {Herbst}, E., {Loison}, J.-C., {et~al.} 2012, \apjs, 199, 21

\bibitem[{{Weselak} {et~al.}(2010){Weselak}, {Galazutdinov}, {Beletsky}, \&
  {Kre{\l}owski}}]{weselak:10}
{Weselak}, T., {Galazutdinov}, G.~A., {Beletsky}, Y., \& {Kre{\l}owski}, J.
  2010, \mnras, 402, 1991

\bibitem[{{Whiteoak} \& {Gardner}(1970)}]{whiteoak:70}
{Whiteoak}, J.~B. \& {Gardner}, F.~F. 1970, \aplett, 5, 5

\bibitem[{{Wiesemeyer} {et~al.}(2016){Wiesemeyer}, {G{\"u}sten}, {Heyminck},
  {H{\"u}bers}, {Menten}, {Neufeld}, {Richter}, {Simon}, {Stutzki}, {Winkel},
  \& {Wyrowski}}]{wiesemeyer:16}
{Wiesemeyer}, H., {G{\"u}sten}, R., {Heyminck}, S., {et~al.} 2016, \aap, 585,
  A76

\bibitem[{{Wiesemeyer} {et~al.}(2018){Wiesemeyer}, {G{\"u}sten}, {Menten},
  {Dur{\'a}n}, {Csengeri}, {Jacob}, {Simon}, {Stutzki}, \&
  {Wyrowski}}]{wiesemeyer:17}
{Wiesemeyer}, H., {G{\"u}sten}, R., {Menten}, K.~M., {et~al.} 2018, \aap, 612,
  A37

\bibitem[{{Wolfire} {et~al.}(2010){Wolfire}, {Hollenbach}, \&
  {McKee}}]{wolfire:10}
{Wolfire}, M.~G., {Hollenbach}, D., \& {McKee}, C.~F. 2010, \apj, 716, 1191

\bibitem[{{Woon} \& {Herbst}(2009)}]{woon:09}
{Woon}, D.~E. \& {Herbst}, E. 2009, \apjs, 185, 273

\end{thebibliography}

\begin{appendix}

\section{Images}
\label{sec:images}

  \begin{figure}
  \centering
 \includegraphics[width=0.45\textwidth]{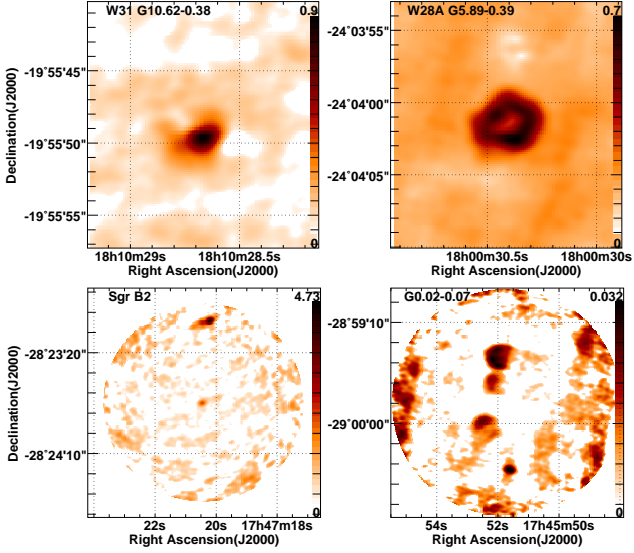}
     \caption{Continuum images at 102 GHz obtained with ALMA. The color bar shows the intensity
     scale in Jy/beam for W31C(G10.6-0.4), W28A2 (G5.89-0.4), SgrB2 and G0.02-0.07 with the peak value given at the top of the color bar. In the SgrB2 image, the source in
     the middle is the SgrB2-S HII region and the strong source at the top is SgrB2-M.
           }
         \label{fig:app:images}
   \end{figure}

Figure \ref{fig:app:images} presents the continuum images at 102~GHz obtained with ALMA.

\section{Spectra}

 \begin{figure*}
 \includegraphics[width=16cm]{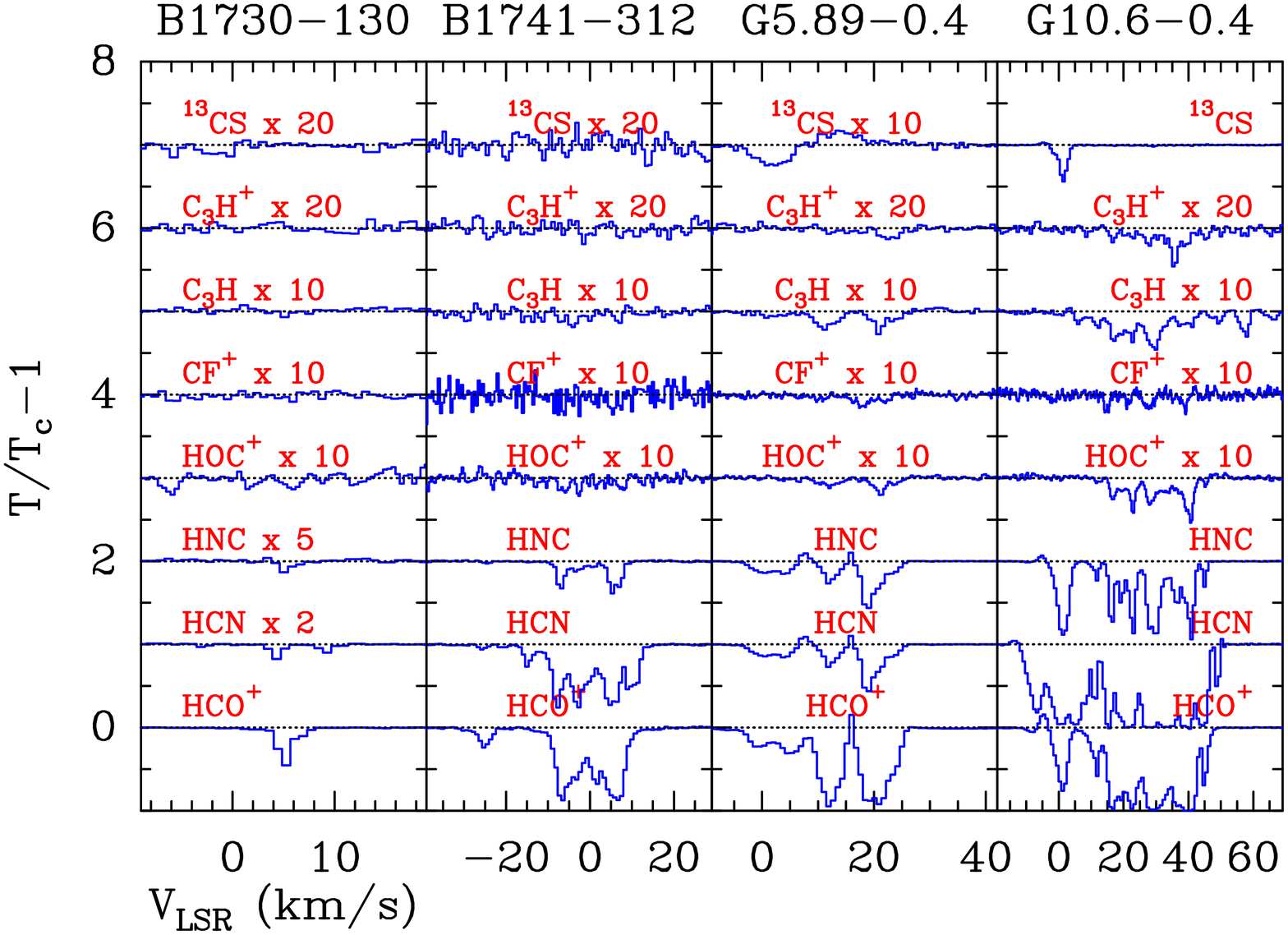}
\vspace*{0.5cm}
  \includegraphics[width=16cm]{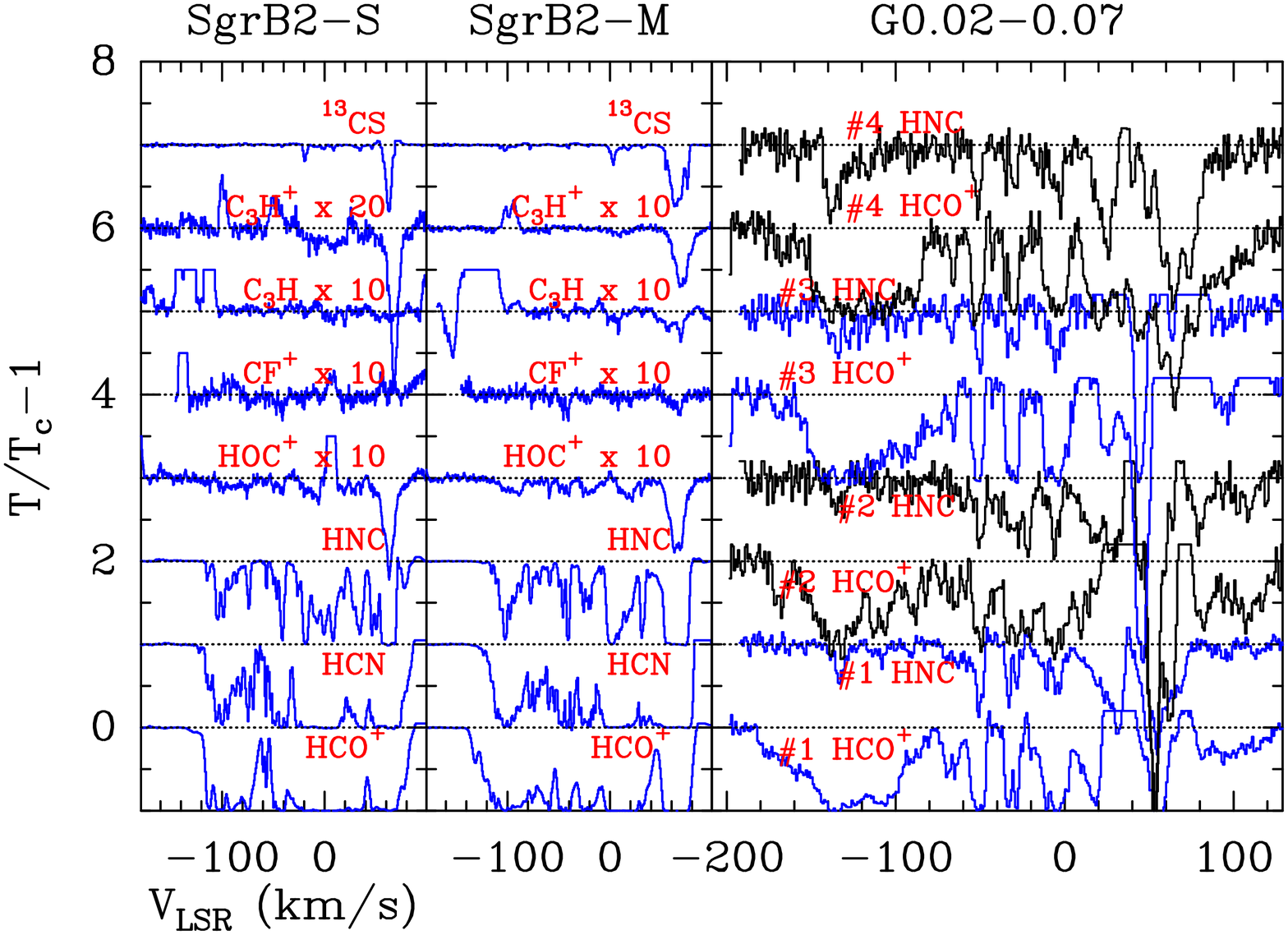}
     \caption{Absorption spectra obtained with ALMA. The spectra are plotted as baseline subtracted line/continuum data. They have been 
     vertically shifted and clipped for clarity. Toward G0.02-0.07 data are displayed toward four positions that correspond to
     the brightest continuum peaks: \#1 at 17:45:51.9, $-$28:59:27.3; \#2 at 17:45:52.1, $-$28:59:40.8; 
     \#3 at 17:45:52.4, $-$29:00:03.5; \#4 at 17:45:51.6, $-$29:00:22.8.
              }
       \label{fig:app-spec}
   \end{figure*}

Figure \ref{fig:app-spec} presents the whole set of absorption spectra. 

\section{Chemistry}
\label{app:chem}

This section describes the analytical formulae used to compute steady state abundances for HOC$^+$ and CF$^+$.
The main formation and destruction routes for HOC$^+$ and CF$^+$ are listed in Table \ref{tab:rates} in which we quote the rates
for HOC$^+$ production, i.e. including the branching ratios.  The reaction rate coefficients  have been taken from the online  database KIDA\footnote{http://kida.obs.u-bordeaux1.fr/} \citep{wakelam:12} and completed by
additional references from the literature as indicated in  Table \ref{tab:rates}.

\begin{table*}
\caption{Main reactions considered}
\label{tab:rates}
\begin{tabular*}{17cm}{lcl}
\hline
 Reaction &  Rate  coefficient (cm$^3$s$^{-1}$ ) & Reference \& Comment \\
\hline
R1 : C$^+$ + H$_2$O  $\rightarrow$ HCO$^+$ + H ; HOC$^+$+H & $k_1$ =  $k_L (0.4767\frac{\mu_D}{\sqrt{2\alpha k_B T}} + 0.62)$& (1)(2)  HOC$^+$ preferred, $k_L =1.04 \times 10^{-9}$ cm$^3$s$^{-1}$\\
R2 : C$^+$ + OH $\rightarrow$ CO$^+$ + H  & $k_2$ =  $k_L (0.4767\frac{\mu_D}{\sqrt{2\alpha k_B T}} + 0.62)$  & (2)(3)  $k_L = 9.1 \times 10^{-10}$ cm$^3$s$^{-1}$\\
R3 : CO$^+$ + H$_2$ $\rightarrow$ HCO$^+$ + H; HOC$^+$ + H & $k_3$ = $7.0 \times 10^{-10}$ & (3)  HOC$^+$ and HCO$^+$ produced equally\\ 
R4 : HOC$^+$ + H$_2$ $\rightarrow$ HCO$^+$ + H$_2$ & $k_4$ = $4.0 \times 10^{-10}$ & (4) \\ 
R5 : C$^+$ + HF $\rightarrow$ CF$^+$ + H                        & $k_5$  =   $7.2 \times 10^{-9} (T/300)^{-0.15}$   &    (5) $k_L = 7.65 \times 10^{-10}$ cm$^3$s$^{-1}$\\                                
R5 : C$^+$ + HF $\rightarrow$ CF$^+$ + H                        & $k_5$  =   $8.64 \times 10^{-10} (T/300)^{-0.43}e^{-12.64/T} $   &    (6) $k_L = 7.65 \times 10^{-10}$ cm$^3$s$^{-1}$ \\                
\hline
R6 : HOC$^+$ + e$^-$ $\rightarrow$ CO + H & $k_e$ = $2.0 \times 10^{-7} (T/300)^{-0.75}$ & (3)  \\
R7 : HCO$^+$ + e$^-$ $\rightarrow$ CO + H & $k"_e$ = $2.8 \times 10^{-7} (T/300)^{-0.69}$ & (7)  \\
R8 : CO$^+$ + e$^-$ $\rightarrow$ C + O & $k'_e$ = $2.75 \times 10^{-7} (T/300)^{-0.55}$ & (3)  \\
R9 : CF$^+$ + e$^-$ $\rightarrow$ C + F &  $kr_e$   = $ 5.2 \times 10^{-8} (T/300)^{-0.8}$  & (8)\\
\hline
\end{tabular*}
\tablefoot{
(1)   \citet{martinez:08} measured $k_1$ =  $2.1 \times 10^{-9}$~cm$^3$s$^{-1}$  at 300~K, (2)\citet{woon:09}, (3) KIDA, (4) \citet{smith:02}, (5) \cite{nw:09}, (6) \cite{denis-alpizar:18}  the expression is valid for 50~\K \, $\le T \le$~2000~\K, (7) \cite{hamberg:14}, (8) \cite{novotny:05}.  
}
\end{table*}

As presented by \citet{liszt:04}, the main formation reactions for HOC$^+$ is reaction R1, with a secondary contribution from reaction R3. For destruction, we
only consider the dissociative recombination with electrons (Reaction R6) and the isomerization reaction with H$_2$ which transforms HOC$^+$ in HCO$^+$ (Reaction R4).
We can therefore write the HOC$^+$ formation rate as :
\begin{equation}
\frac{dn(HOC^+)}{dt}_{form} = k_1n(C^+)n(H_2O) + k_3n(CO^+)n(H_2)
\end{equation}

and the destruction rate as :
\begin{equation}
\frac{dn(HOC^+)}{dt}_{dest} = k_en(HOC^+)n(e^-) + k_4n(HOC^+)n(H_2) .
\end{equation}

We further assume that C$^+$ is the main carrier of carbon and all electrons are provided by C$^+$, which leads to:
\begin{equation}
n(C^+) = n(e^-) = x_C n_H = 2 x_C n(H_2)/f(H_2) ,
\end{equation}

where we have introduced the fraction of hydrogen in molecular form, $f(H_2) = \frac{2n(H_2)}{n(HI)+2n(H_2)}$, and $n_H = n(HI)+2n(H_2)$, and the carbon
abundance relative to H, $x_C$.

Equating the formation and destruction rates for HOC$^+$, and writing $n(H_2O) = [H_2O]n(H_2)$, where $[H_2O]$ represents the abundance of water relative to H$_2$,
 we get :
\begin{equation}
[HOC^+] = \frac{n(HOC^+)}{n(H_2)} = \frac{2 k_1 x_C [H_2O] + k_3[CO^+]f(H_2)}{2k_e x_C + k_4f(H_2)} .
\end{equation}

CO$^+$ is formed in reaction R2 and destroyed by reacting with electrons (R8) and with H$_2$ (R3). We therefore obtain the CO$^+$ abundance relative to H$_2$, $[CO^+]$,
 in a similar way, where $[OH]$ represents the abundance of OH relative to H$_2$ :
 
\begin{equation}
[CO^+] = \frac{ k_2  x_C [OH] }{ k'_e x_C  + k_3 f(H_2)} .
\end{equation}

By substituting [CO$^+$], we obtain an equation for the ratio of the HOC$^+$ and H$_2$O abundances :

\begin{equation}
\frac{[HOC^+]}{[H_2O]} = x_C \frac{2 k_1(k'_e x_C + k_3 f(H_2))  + k_3 k_2 f(H_2)\frac{ [OH]}{[H_2O]}}{(k'_e x_C  + k_3 f(H_2) s)(2k_e x_C + k_4f(H_2))} .
\end{equation}

We can also predict the ratio of HCO$^+$ and H$_2$O abundances using the same chemical network, noting $\alpha$ the small branching
ratio toward HCO$^+$ in reaction R1 :

\begin{equation}
\frac{[HCO^+]}{[H_2O]} =   \\
\frac{2k_1 \alpha (k'_e x_C +k_3 f(H_2)) + k_3k_2f(H_2) \frac{[OH]}{[H_2O]} + k_4 f(H_2) \frac{[HOC^+]}{[H_2O]}}{2 k"_e x_C (k'_e x_C + k_3 f(H_2))} .
\end{equation}

\begin{figure*}
\centering
 \includegraphics[width=0.45\textwidth]{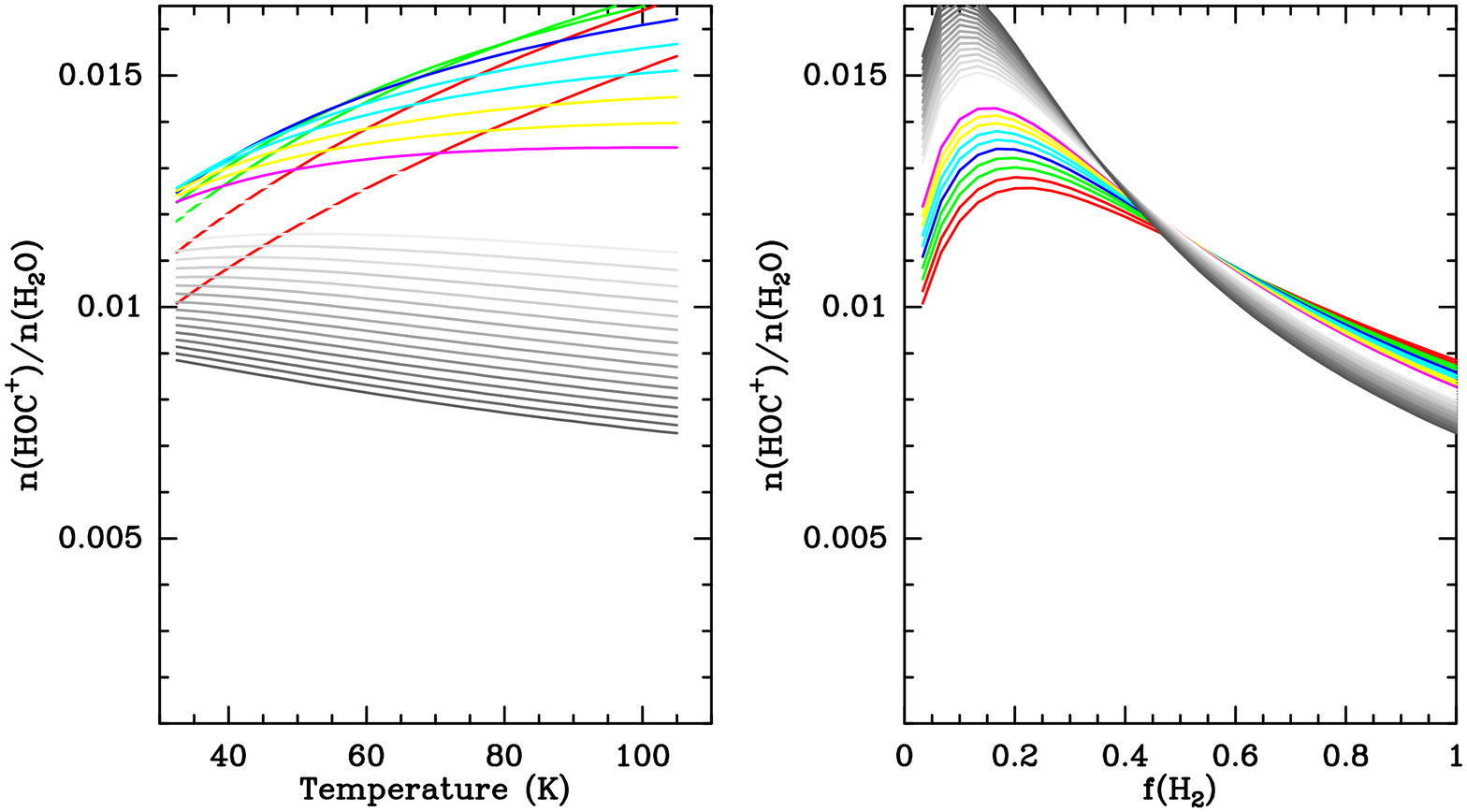}
  \includegraphics[width=0.45\textwidth]{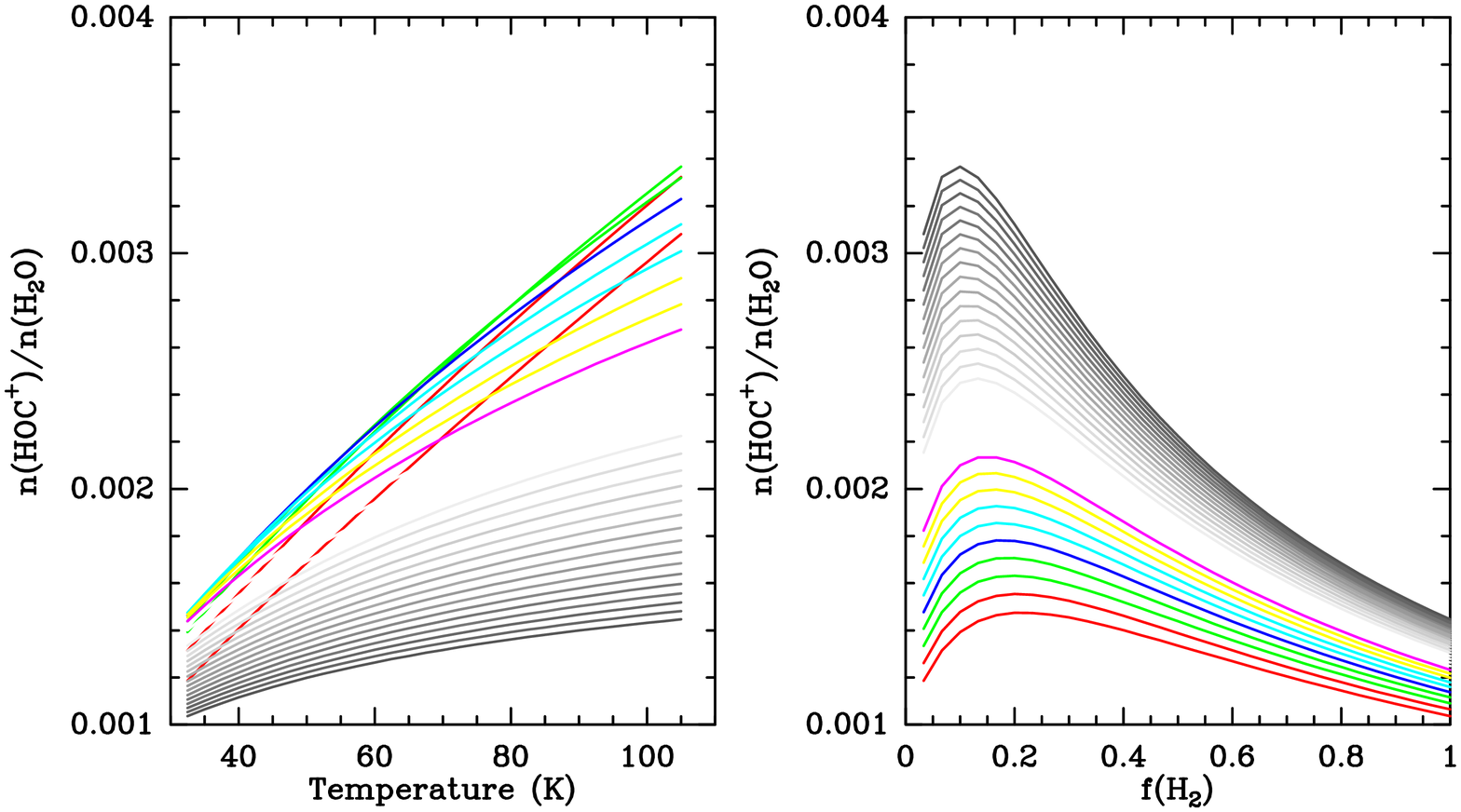}
\caption{\label{fig:rates} Left : Variation of the ratio of the HOC$^+$ and H$_2$O abundance with the kinetic temperature. Each curve
represents a different fraction of hydrogen in molecular form f(H$_2)$ from 0.03 (red) up to 1 (black). 
Right : Variation of the ratio of the HOC$^+$ and H$_2$O abundance with f(H$_2)$. Each curve
represents a different  temperature  from 32~K (red) up to 105~K (black). The two panels on the left use  enhanced capture rates
for reactions between ions and polar molecules using the formalism and the molecular data described by \citet{woon:09}. The right panels
use the Langevin rates computed from the molecular data listed by \citet{woon:09}. The observed N(HOC$^+$)/N(H$_2$O) ratio is
$1.8 \times 10^{-3}$, see Table \ref{tab:ratio}. }
\end{figure*}

\begin{figure*}
\centering
 \includegraphics[width=0.8\textwidth]{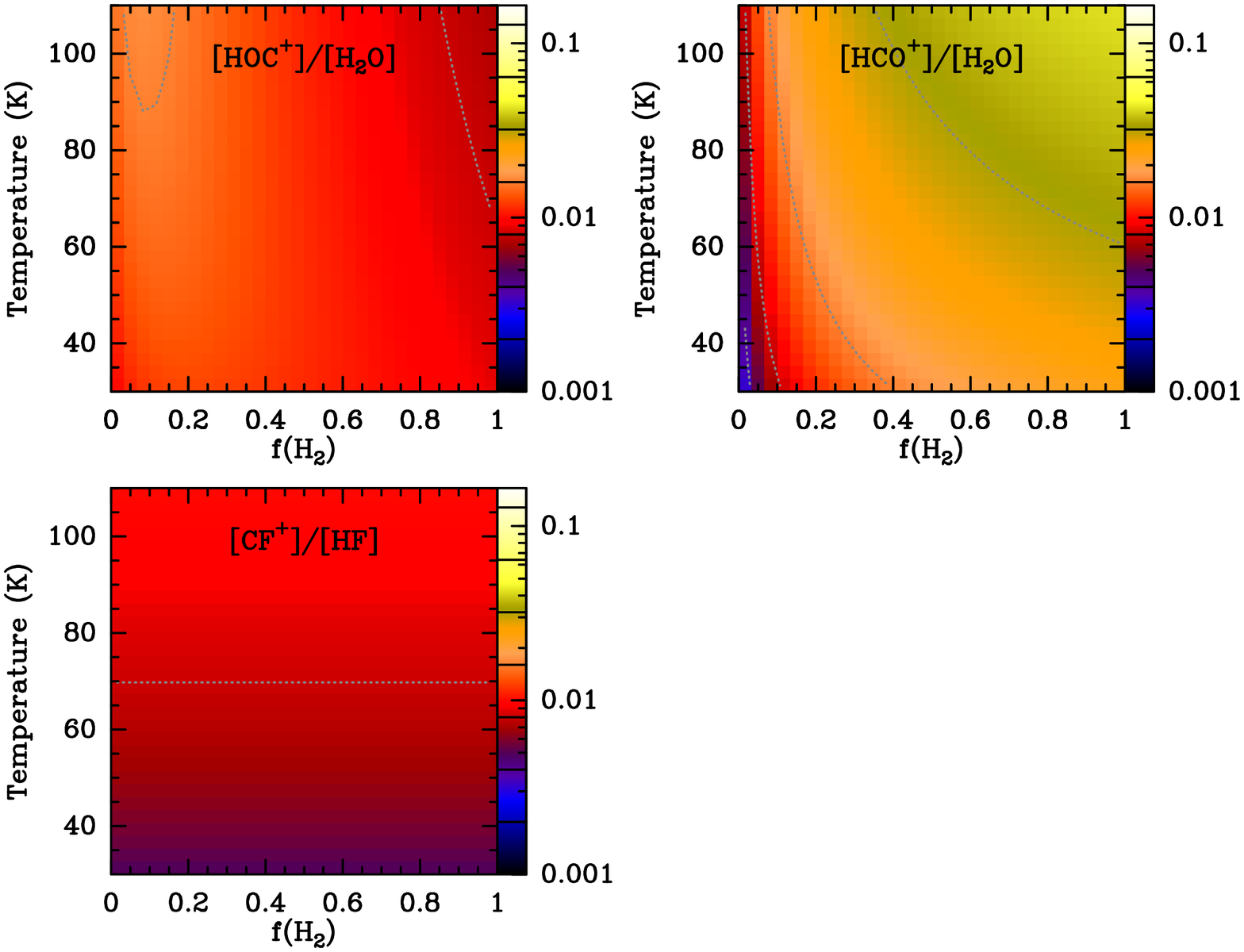}
\caption{\label{fig:rates1}  Variation of the predicted abundance ratios [HOC$^+$]/[H$_2$O], [HCO$^+$]/[H$_2$O], and [CF$^+$]/[HF] as a function of
the fraction of molecular gas f(H$_2$) and the kinetic temperature. The plot is using enhanced capture rates for reactions between
ions and polar molecules using the formalism and the molecular data described by \citet{woon:09} and the new theoretical reaction rate for reaction R5 \citep{denis-alpizar:18}.  Contour levels are set at $2\times 10^{-3}$,
 $4\times 10^{-3}$,  $8\times 10^{-3}$, ...up to $128 \times 10^{-3}$. The mean values of the abundance ratios are listed in Table~\ref{tab:ratio}.}
\end{figure*}

Figures \ref{fig:rates} and \ref{fig:rates1} illustrate the expected variation of the ratio of the HOC$^+$ and H$_2$O abundances with the kinetic temperature and with the fraction of hydrogen
in molecular form $f(H_2)$. We used two assumptions for reactions R1, R2 and R5 : the pure capture model (the Langevin rate), and the ion-molecule
capture rate which leads to higher value and a strong temperature dependence \citep{woon:09}. 
In this approach, the reaction rate can be written: $k = k_L (0.4767 x + 0.62)$, where $x = \frac{\mu_D}{\sqrt{2\alpha k_B T}}$, $\mu_D$ is the dipole moment of the polar molecule, $\alpha$ is
the polarizability, and $T$ the kinetic temperature. The formula is valid for $x  \ge 2$, which is the case for the reactions considered in this work.

\citet{martinez:08} have 
studied the C$^+$ + H$_2$O reaction experimentally at 300~K using a flowing afterglow-selected ion flow tube (FASIFT). They determine
a total rate of $2.1 \times 10^{-9}$~cm$^{3}$s$^{-1}$, a factor of two higher than the Langevin rate. The branching ratios between
HCO$^+$ and HOC$^+$ are unknown.

\begin{figure*}
\centering
 \includegraphics[width=0.8\textwidth]{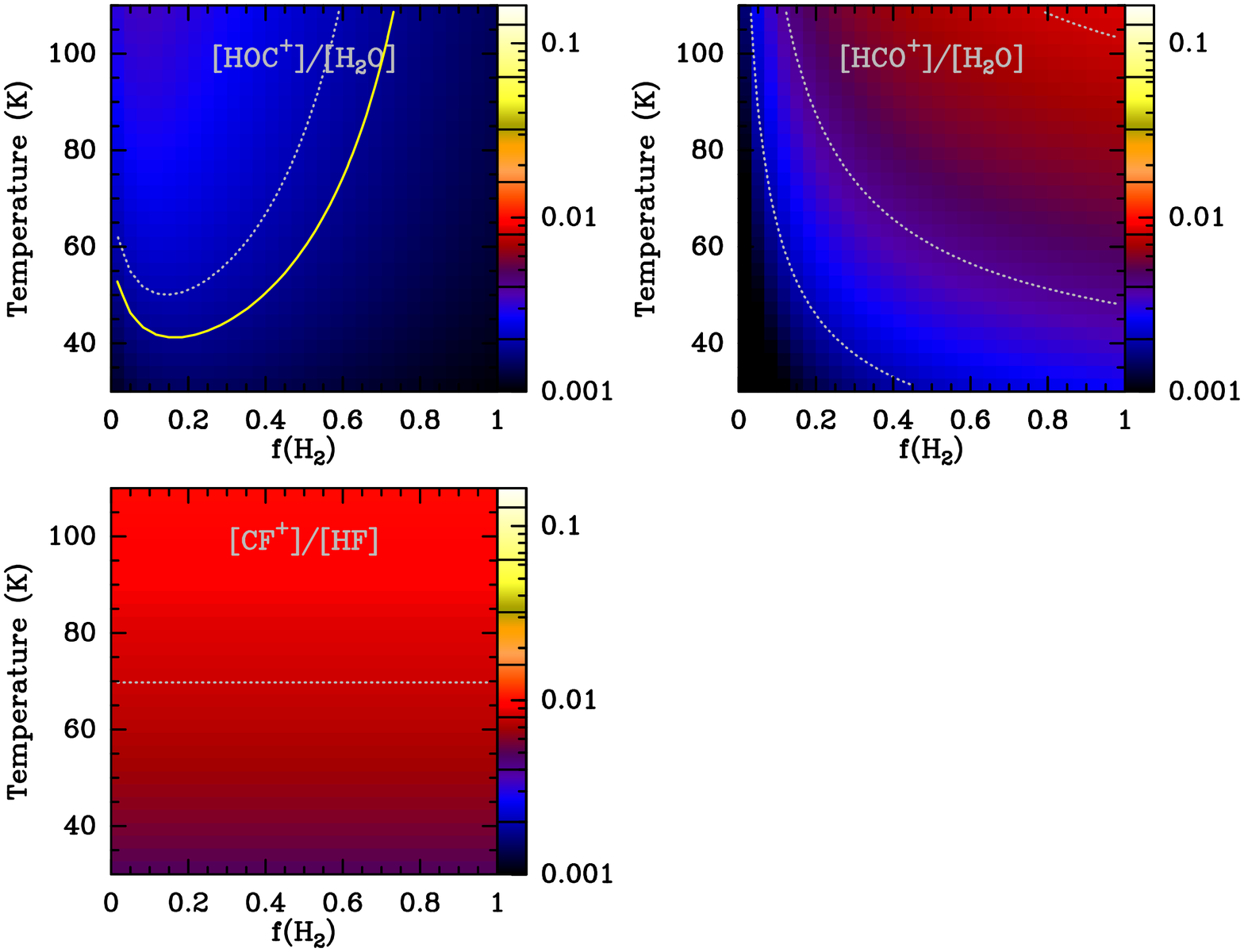}
\caption{\label{fig:rates2}
Variation of the predicted abundance ratios [HOC$^+$]/[H$_2$O], [HCO$^+$]/[H$_2$O], and [CF$^+$]/[HF] as functions of
the fraction of molecular gas f(H$_2$) and the kinetic temperature. The plot uses Langevin  rates for reactions between
ions and polar molecules with the molecular data listed by \citet{woon:09} and the new theoretical reaction rate for reaction R5 \citep{denis-alpizar:18}. Contour levels are set at $2\times 10^{-3}$,
 $4\times 10^{-3}$,  $8\times 10^{-3}$, ...up to $128 \times 10^{-3}$. The yellow contour in the top left box indicates the mean observed ratio [HOC$^+$]/[H$_2$O] $= 1.8\times 10^{-3}$. The mean values of the abundance ratios are listed in Table~\ref{tab:ratio}. }
\end{figure*}

\end{appendix}

\end{document}